\documentclass[aps,prd,groupedaddress,preprintnumbers,%
              showpacs,showkeys,floatfix]{revtex4-1}       % Documentstyle for
\usepackage[utf8]{inputenc}                    % Codepage latin1
\usepackage{graphicx}                            % Graphics package
\usepackage{epsfig}
\usepackage{epsf}
\usepackage{latexsym}                            % Load additional symbols
\usepackage{amsfonts}                            % Use AMS fonts,
\usepackage{amssymb}                             % symbols,
\usepackage{amsmath}                             % and definitions
\usepackage[mathscr]{eucal}                      % Euler font symbols
\usepackage{dcolumn}                             % Align table columns
\usepackage{multirow}
\usepackage{bm}                                  % Bold math
\usepackage{hyperref}                            % References as links
\usepackage[usenames]{color}
\usepackage{array}
\usepackage{appendix}
\usepackage{bbold}
\usepackage{mathtools}
\graphicspath{{.}{Graphics/}}                    % Path for graphics
\newcommand{\be}{\begin{equation}}
\newcommand{\ee}{\end{equation}}
\newcommand{\bea}{\begin{eqnarray}}
\newcommand{\eea}{\end{eqnarray}}

\def\eq#1{Eq.~(\ref{#1})}
\def\fig#1{Fig. \ref{#1}}
\def\tbl#1{Table \ref{#1}}
\def \3{\ss }

\newcommand{\tr}{\operatorname{Tr}}

\newcommand{\<}{\langle}
\renewcommand{\>}{\rangle}

\def\cyp{a}
\def\cyi{b}
\def\temple{c}
\def\itad{d}
\def\ita{e}
\def\desy{f}
\def\bonn{g}
\def\gre{h}
\def\utah{i}
\def\bern{j}

\begin{document}

\begin{titlepage}
  {\vspace{-0.5cm} \normalsize
  \hfill \parbox{60mm}{%DESY 14-096\\
                       %LPT-ORSAY 08-32 \\ 
                       %IRFU-08-29\\ 
                       % ROM2F/2008/06\\
                       %    HU-EP-08/09\\
                       %     MS-TP-08-4.\\
                        %RM3-TH/, ROM2F/2007/\\
}}\\[10mm]
\begin{center}
\begin{LARGE}
\textbf{Nucleon scalar and tensor charges using lattice QCD simulations at the physical value of the pion mass} \\
\end{LARGE}
\end{center}

\vspace{-0.3cm}
\baselineskip 20pt plus 2pt minus 2pt
\begin{center}
\textbf{
      C.~Alexandrou$^{(\cyp, \cyi)}$,
	  M. Constantinou$^{(\temple)}$,
      P. Dimopoulos$^{(\itad, \ita)}$,
        R. Frezzotti$^{(\ita)}$,
	  K.~Hadjiyiannakou$^{(\cyi)}$,
	  K.~Jansen$^{(\desy)}$,
      C.~Kallidonis$^{(\cyi)}$, 
       B.~Kostrzewa$^{(\bonn)}$,
	  G.~Koutsou$^{(\cyi)}$,
      M.~Mangin-Brinet$^{(\gre)}$,
 	  A.~Vaquero Avil\`es-Casco$^{(\utah)}$,
      U. Wenger$^{(\bern)}$
}
\end{center}

\begin{center}
\begin{footnotesize}
\noindent 	
 	$^{(\cyp)}$ Department of Physics, University of Cyprus, P.O. Box 20537,
 	1678 Nicosia, Cyprus\\	
 	$^{(\cyi)}$ Computation-based Science and Technology Research Center, The Cyprus Institute, 20 Kavafi Str., Nicosia 2121, Cyprus \\
$^{(\temple)}$ Physics Department, Temple University, 1925 N. 12th Street,
Philadelphia, PA 19122-1801\\
$^{(\itad)}$ Centro Fermi - Museo Storico della Fisica e Centro Studi e Ricerche ``Enrico Fermi", Compendio del Viminale, Piazza del Viminiale 1, I-00184, Rome, Italy\\
$^{(\ita)}$ Dipartimento di Fisica, Universit\'a di Roma Tor Vergata, Via della Ricerca Scientifica 1, I-00133 Rome, Italy\\
%$^{(\roma)}$ Dipartimento di Fisica, Universit\'a and INFN di Roma Tor Vergata, 00133 Roma, Italy\\
	$^{(\desy)}$ NIC, DESY, Platanenallee 6, D-15738 Zeuthen, Germany\\
$^{(\bonn)}$ HISKP (Theory), Rheinische Friedrich-Wilhelms Universit\"at Bonn, 53115 Bonn, Germany\\
$^{(\gre)}$ Theory Group, Lab. de Physique Subatomique et de Cosmologie, 38026 Grenoble, France\\
    $^{(\utah)}$ Department of Physics and Astronomy, University of Utah, Salt Lake City, UT 84112, USA\\
$^{(\bern)}$ Albert Einstein Center for Fundamental Physics, University of Bern, 3012 Bern, Switzerland

\vspace{0.2cm}
\end{footnotesize}
\end{center}

\begin{abstract}

We present results on the light, strange and charm nucleon scalar and tensor charges from lattice QCD, using simulations with $N_f=2$ flavors of twisted mass Clover-improved fermions with a physical value of the pion mass.  Both connected and disconnected contributions are included, enabling us to extract the isoscalar, strange and charm charges for the first time directly at the physical point. 
Furthermore, the renormalization is computed  non-perturbatively for both isovector and isoscalar quantities. We investigate excited state effects by analyzing several sink-source time separations and by employing a set of methods to probe ground state dominance.  Our final results for the scalar charges are $g_S^u = 5.20(42)(15)(12)$, $g_S^d = 4.27(26)(15)(12)$, $g_S^s=0.33(7)(1)(4)$, $g_S^c=0.062(13)(3)(5)$ and for the tensor charges $g_T^u = 0.782(16)(2)(13)$, $g_T^d =  -0.219(10)(2)(13)$, $g_T^s=-0.00319(69)(2)(22)$, $g_T^c=-0.00263(269)(2)(37)$ in the $\overline{\rm MS}$ scheme at 2~GeV. The first error is statistical, the second is the systematic error due to  the renormalization and the third the systematic arising from possible contamination due to the excited states. 
\begin{center}
\today
\end{center}

\end{abstract}

\pacs{11.15.Ha, 12.38.Gc, 12.38.Aw, 12.38.-t, 14.70.Dj}
\keywords{Lattice QCD, scalar charge, tensor charge}
\maketitle 

\end{titlepage}

%\tableofcontents
\section{Introduction}

The nucleon scalar and tensor charges are fundamental properties of hadron structure but most importantly they  are related to the ongoing search for new physics beyond the Standard Model (BSM). The nucleon isovector scalar and tensor charges  probe novel scalar and tensor interactions at the TeV scale. Planned neutron $\beta$-decay experiments with higher accuracy would require input on the scalar and tensor charges. 
Furthermore, the nucleon matrix element of the light, strange and charm scalar quark operator, from which the scalar charge is extracted, is directly related to the nucleon scalar contents or $\sigma$-terms. These quantities are crucial input  in experimental  dark matter searches~\cite{Cushman:2013zza} that are seeking to directly detect dark matter by measuring the recoil energy of scattering between nuclei and dark matter candidates. These candidates are weakly-interacting massive particles (WIMPs) and according to a number of BSM theories~\cite{Ellis:2010kf,Servant:2002hb,Bertone:2010ww,Bertone:2004pz} they interact with normal matter via elastic scattering. During the scattering process, a WIMP produces a Higgs boson, which then interacts with a nucleon through scalar density operators. For spin-independent elastic scattering, the theoretical expression of the cross section depends quadratically on the nucleon scalar matrix element. This contribution, in fact, brings the largest uncertainty on the nucleon dark matter cross section~\cite{Ellis:2008hf}. 

The nucleon tensor charge plays an important role 
in BSM physics connected to novel CP-violating interactions. Such interactions   will lead to a non-zero neutron electric dipole moment (nEDM) and planned experiments to reduce the current bound by two orders of magnitude will constrain many BSM theories. An accurate measurement of the flavor-diagonal tensor charges will be needed in order to translate the new bounds on the nEDM into CP-violating terms in BSM theories and set bounds on new sources of CP-violation~\cite{Bhattacharya:2016zcn}. Accurate values of both scalar and tensor charges are needed also in evaluating
the contribution of the CP-odd electron-nucleon interaction contributing to the atomic EDM~\cite{Yamanaka:2017mef}.

Unlike the axial charge, $g_A$, the scalar and tensor charges are not well known experimentally. The $0^+ \rightarrow 0^+$ nuclear decays and the radiative pion decay $\pi \rightarrow e\nu\gamma$, respectively, provide limits on their values. Experiments using  ultra-cold neutrons are  expected to improve these values~\cite{Bhattacharya:2011qm}. In addition, there is a rich experimental program to study the transverse spin structure of the nucleon at Jefferson Lab.  A coincidence experiment in Hall A will employ a newly proposed solenoid spectrometer (SoLID) to perform precision measurements from semi-inclusive electro-production of charged pions from  transversely polarized $^3$He target in Deep-Inelastic-Scattering kinematics using 11 and 8.8 GeV electron beams~\cite{Gao:2010av}.
 SoLID is expected to increase the experimental accuracy of the tensor coupling by an order of magnitude~\cite{Dudek:2012vr,Ye:2016prn}. On-going experiments at LHC are also probing scalar and tensor interactions for BSM physics   at the TeV scale, and  they are expected to increase the limits to contributions arising from such interactions by an order of magnitude. This experimental activity makes a precise lattice QCD calculation of the scalar and tensor couplings well-timed: It  provides valuable input in the ongoing searches for BSM physics, and sheds light on our understanding of nucleon structure. 

Lattice QCD has progressed noticeably in the last few years,  due to new algorithmic improvements and the increase in the available computational power. These ongoing advancements allow for lattice QCD simulations at physical values of the pion mass and  at increasingly larger  volumes. Such simulations eliminate the need for chiral extrapolations, thus reducing a significant source of systematic uncertainties. However, calculations of baryon observables close to or at the physical point have a worse  signal-to-noise ratio and larger effects due to 
 excited-state contaminations, rendering such calculations more challenging. Typically one needs one order of magnitude larger statistics as compared to using simulations with heavier pion masses for the same setup. To eliminate excited states one needs larger Euclidean time propagation with exponentially increasing statistical noise and thus large statistics.

In this work we study the light, strange and charm scalar and tensor nucleon charges using a gauge ensemble with two degenerate light flavors ($N_f=2$) of twisted mass clover-improved fermions with pion mass fixed to its physical value~\cite{Abdel-Rehim:2015pwa}.  Since we are analyzing a single gauge ensemble, cut-off and finite volume effects cannot be evaluated using directly lattice results. 

The isovector scalar and tensor charges,  $g_S^{u-d}$ and $g_T^{u-d}$, are straight forward to calculate since they receive only connected contributions arising from the coupling of the operator to valence quarks, as depicted in Fig~\ref{fig:3pt_diagrams}. Several lattice QCD results have been obtained recently including direct evaluation at the physical point~\cite{Abdel-Rehim:2015owa,Bhattacharya:2016zcn}. The  isoscalar charges $g_S^{u+d}$ and $g_T^{u+d}$ receive additional contributions coming from the coupling of the operator to vacuum quarks, forming disconnected quark loops. The strange and charm charges $g_S^{s,c}$ and $g_T^{s,c}$ receive purely disconnected contributions which are  notoriously difficult to evaluate being computationally very demanding.  It is only recently that disconnected diagrams were included in lattice QCD calculations of the scalar and tensor matrix elements~\cite{Bali:2011ks,Freeman:2012ry,Gong:2013vja,Abdel-Rehim:2013wlz,Bhattacharya:2015wna,Yamanaka:2015lfk,Abdel-Rehim:2016won} eliminating an uncontrolled systematic uncertainty. In this work we employ improved stochastic methods to include all the disconnected diagrams with satisfactory accuracy and by applying non-perturbative renormalization to  obtain results on the strange and charm scalar and tensor charges with all contributions taken into account directly at the physical point mass.

The paper is organised as follows: In section~\ref{sec:simulation details} we summarize the characteristics of the gauge configurations used and  in section~\ref{sec:matrix elements} we describe the extraction of the appropriate matrix elements, which for zero momentum transfer yield the charges. In section~\ref{sec:lattice evaluation} we discuss in detail the lattice QCD computation of both connected and disconnected contributions and their renormalization, and in section~\ref{sec:results} we present our results. In section~\ref{sec:comparisons} we compare our results with those obtained recently by other lattice QCD groups and in section~\ref{sec:conclusions} we conclude.

%==========================================================================
%==========================================================================

\section{Simulation details}
\label{sec:simulation details}

We analyze an $N_f=2$ gauge ensemble produced by the European Twisted Mass Collaboration (ETMC)~\cite{Abdel-Rehim:2015pwa,Abdel-Rehim:2015owa} at the physical pion mass. The ``Iwasaki" improved gauge action~\cite{Iwasaki:1983ck,Abdel-Rehim:2013yaa} is employed. The lattice volume is $48^3\times 96$ and the lattice spacing determined from the nucleon mass is $a=0.0938(2)$~fm. The rest of the parameters regarding this ensemble are listed in~\tbl{Table:sim_params}. We shall refer to this ensemble as the ``physical ensemble" from now on. In the fermion sector, the twisted mass fermion (TMF) action at maximal twist is employed~\cite{Frezzotti:2000nk,Frezzotti:2003ni}, including a Clover term~\cite{Sheikholeslami:1985ij}
\be\label{eq:S_tml}
S_F\left[\chi,\overline{\chi},U \right]= a^4\sum_x  \overline{\chi}(x)\left(D_W[U] + m_{\rm cr} + i \mu_l \gamma_5\tau^3 - \frac{1}{4}c_{\rm SW}\sigma^{\mu\nu}\mathcal{F}^{\mu\nu}[U] \right) \chi(x)\;,
\ee
where $D_W[U]$ denotes the massless Wilson-Dirac operator, $\tau^3$ is the third Pauli matrix acting in flavour space, $m_{\rm cr}$ is the bare untwisted mass tuned to its critical value and $\mu_l$ is the bare twisted light quark mass. The last term in~\eq{eq:S_tml} is the clover-term, with $c_{\rm SW}$ the so-called Sheikoleslami-Wohlert improvement coefficient, which is fixed to $c_{\rm SW}=1.57551$~\cite{Aoki:2005et}, $\mathcal{F}^{\mu\nu}[U]$ is the field strength tensor and $\sigma^{\mu\nu} = (1/2)[\gamma_\mu,\gamma_\nu]$. With $\chi(x)$ we denote the light quark doublet in the twisted basis, $\chi = (u,d)$.

The TMF action is particularly attractive for hadron structure calculations as it provides an automatic $\mathcal{O}(a)$ improvement  without requiring  further operator improvement. Additional advantages are the infrared regularization of small eigenvalues that makes  dynamical simulations faster and the simplified renormalization of operators~\cite{Frezzotti:2003ni,Jansen:2005cg,Farchioni:2004us,Farchioni:2005bh}. However, due to  $\mathcal{O}(a^2)$ lattice artefacts that lead to instabilities in the numerical simulations, particularly at quark masses close to their physical values, the addition of a clover term was required. The latter reduces isospin symmetry breaking effects, while preserving the automatic $\mathcal{O}(a)$ improvement. Another advantage of TMF is that scalar matrix elements are multiplicatively renormalizable~\cite{Dinter:2012tt}, hence a mixing between the bare light and strange scalar matrix elements, seen in other Wilson-type fermion actions, does not occur. TMF also obey a powerful property, which allows an effective increase of the signal-to-noise ratio of the disconnected quark loops, known as the one-end trick~\cite{Michael:2007vn,Foster:1998vw,McNeile:2006bz}. In section~\ref{sec:lattice evaluation} we give more details on the techniques used 
for the computation of the disconnected contributions. The reader interested in more technical details regarding the twisted mass action and the simulations of the gauge ensemble used in this work is referred to Refs.~\cite{Abdel-Rehim:2015pwa,Abdel-Rehim:2014nka,Boucaud:2007uk,Boucaud:2008xu,Baron:2009wt}.
\begin{table}[h]
\begin{center}
\renewcommand{\arraystretch}{1.2}
\renewcommand{\tabcolsep}{5.5pt}
\begin{tabular}{c|lc}
\hline\hline
\multicolumn{3}{c}{ $\beta=2.10$, $a=0.0938(3)(2)$~fm,  ${r_0/a} = 5.32(5)$ }\\
\hline
\multirow{3}{*}{$48^3\times 96$, $L=4.5$~fm}  & $a\mu$         & 0.0009    \\
                              			      & $m_\pi$~(GeV)  & 0.1305(4) \\
                        					  & $m_\pi L$      & 2.98      \\
\hline\hline
\end{tabular}
\caption{Input parameters ($\beta,L, a\mu$) of our lattice simulation with the corresponding lattice spacing and pion mass. The systematic error on the lattice spacing given in the second parenthesis is due to the interpolation to 135~MeV pion mass. The value of the lattice spacing is determined from the nucleon mass  using 140 times the statistics as compared to what was used in Refs.~\cite{Abdel-Rehim:2015owa,Abdel-Rehim:2015pwa}, namely using ${\cal O}(215,000)$ statistics.}
\label{Table:sim_params}
\end{center}
\vspace*{-.0cm}
\end{table} 

%==========================================================================
%==========================================================================

\section{Matrix element decomposition}
\label{sec:matrix elements}

The quantity of interest is the forward nucleon  matrix element $\< N(p)|\mathcal{O}_\Gamma|N(p)\>$, where $|N(p)\>$ is a nucleon state with momentum $p$ and $\mathcal{O}_\Gamma$ is either the local scalar or tensor operator. In the physical basis, these operators read
\be\label{eq:operators}
\mathcal{O}_{S^a} = \bar{q}\frac{\tau^a}{2}q\;,\quad \mathcal{O}_{T^a}^{\mu\nu} = \bar{q}\sigma^{\mu\nu}\frac{\tau^a}{2}q\;,
\ee
respectively, where $q = u,d$  and $\sigma^{\mu\nu} = (1/2)[\gamma_\mu,\gamma_\nu]$. The $\tau^a$ matrix acts in flavour space. We consider both isovector and isoscalar quantities, for which we take $\tau^a = \tau^3$ and $\tau^a = \mathbb{1}$, respectively.   The individual contributions for $g_S^{u,d}$ and $g_T^{u,d}$ can then be extracted from the isovector and isoscalar combinations. This is equivalent to calculating directly these contributions by substituting $\tau^a$ with the corresponding projectors onto the up- and down-quarks in~\eq{eq:operators}. Unless otherwise specified, all expressions are given in Euclidean space. For the  strange and charm quarks we use Osterwalder-Seiler fermions, that is, they are introduced as heavy doublets similar to the light quark doublet, $\chi(f) = (f^+,f^-)$, where $f = s,c$. The action for these doublets is the same as~\eq{eq:S_tml}, but with the light twisted mass $\mu_l$ replaced by the corresponding mass of the given heavy quark, $\mu_f$, and $f^\pm$ refers to choosing $\pm \mu_f$. We have tuned the bare twisted mass of the strange and charm quarks to reproduce the physical $\Omega^-$ and $\Lambda_c^+$ mass, respectively. The values we obtain are $a\mu_s = 0.0259(3)$ and $a\mu_c = 0.3319(15)$~\cite{Alexandrou:2017xwd}. At our fixed lattice spacing we get for the renormalized quark masses at 2~GeV in the $\overline{\rm MS}$-scheme are

\be
m_s^R=108.6(2.2)~{\rm MeV}\;,\; m_c^R=1392.6(23.5)~{\rm MeV},
\label{quark mass}
\ee
where only statistical errors are quoted. A more complete analysis, including systematic errors will follow in the future.
The nucleon scalar and tensor charges can be extracted from the corresponding matrix elements of the operators of~\eq{eq:operators} at zero momentum transfer, which are decomposed as
\bea \label{eq:scalar_decomp}
\<N(p,s')|\mathcal{O}_{S^a}|N(p,s)\> &=& \bar{u}_N(p,s')\left[\frac{1}{2}G^a_S(0)\right]u_N(p,s)\quad , \\  \label{eq:tensor_decomp}
\<N(p,s')|\mathcal{O}_{T^a|}N(p,s)\> &=& \bar{u}_N(p,s')\left[\frac{1}{2}A^a_{T10}(0)\sigma^{\mu\nu}\right]u_N(p,s) \;.
\eea
From the above matrix elements, the scalar and tensor charges can be obtained from $G_S(0)\equiv g_S $ and $A_{T10}(0)\equiv g_T$. Depending on whether the operators are either the individual up- or down-quark contributions or the isovector or isoscalar combinations, the corresponding charge is obtained. We note here that for non-zero momentum transfer,  the form factors $B_{T10}(Q^2)$ and $\tilde{A}_{T10}(Q^2)$ appear in~\eq{eq:tensor_decomp}, where $Q^2$ is the momentum transfer in Euclidean space. We do not consider these form factors in this work.

%==========================================================================
%==========================================================================

\section{Lattice evaluation}
\label{sec:lattice evaluation}

\subsection{Correlation functions}\label{sec:corr_funcs}

In lattice QCD,  the matrix elements of Eqs.~(\ref{eq:scalar_decomp}) and~(\ref{eq:tensor_decomp}) are computed by  constructing appropriate  three-point correlation functions. Since we are interested in extracting the charges we need the matrix elements with zero momentum transfer. We thus give here the corresponding expression for the three-point function for the case $\vec{q}=\vec{0}$ as well as the nucleon two-point function needed for canceling the Euclidean time evolution and unknown overlaps of the interpolating field with the nucleon state: 
\bea
\label{eq:3ptfunc}
G_\Gamma^{\rm 3pt}(P,\vec{p},\vec{q}=\vec{0},t_s,t_{\rm ins}) &=& \sum_{\vec{x}_s,\vec{x}_{\rm ins}} e^{-i(\vec{x}_s - \vec{x}_0)\cdot \vec{p}} P_{\beta\alpha}\<J_\alpha (\vec{x}_s,t_s)\mathcal{O}_\Gamma(x_{\rm ins},t_{\rm ins})\bar{J}_\beta(\vec{x}_0,t_0)\>\;, \\ 
 \label{eq:2ptfunc}
G^{\rm 2pt}(\vec{p},t_s) &=& \sum_{\vec{x}_s}P^4_{\beta\alpha} \<J_\alpha (\vec{x}_s,t_s)\bar{J}_\beta(\vec{x}_0,t_0)\> e^{-i\vec{x}_s \cdot \vec{p}}\;,
\eea
where $t_0$, $t_{\rm ins}$ and $t_s$ are the source, insertion and sink time coordinates, respectively. The projector matrix $P^4$ is given by
\be\label{eq:proj_4}
P^4 = \frac{1}{4}\left(\mathbb{1}\pm\gamma_0\right)\;.
\ee
For the scalar charge, the unpolarized projector $P = P^4$ is used in the three-point function, whereas for the tensor charge, the polarized projector
\be\label{eq:proj_pol}
P = P_k \equiv iP^4\gamma_5\gamma_k\;, k=1,2,3
\ee
is required. We work in the rest frame, i.e. the source and sink carry zero momentum, therefore we also set $\vec{p}=0$. We use the proton interpolating operator given by
\be\label{eq:proton_intfield}
J_\alpha(x) = \epsilon^{abc}u_\alpha^a \left[u_\beta^b(x)(C\gamma_5)_{\beta\gamma}d_\gamma^c(x)\right]\quad.
\ee
In order to increase the overlap with the proton ground state we apply Gaussian smearing~\cite{Alexandrou:1992ti,Gusken:1989ad} at the source and sink. The smeared quark fields read
\be\label{eq:Gsmear}
q_{\rm smear}^a(\vec{x},t) = \sum_y F^{ab}(\vec{x},\vec{y};U(t))q^b(\vec{y},t)\;, \quad F(\vec{x},\vec{y};U(t)) = \left(\mathbb{1}+\alpha_G H\right)^{n_G}(\vec{x},\vec{y};U(t))
\ee
and $H$ is the hopping term realized as a matrix in coordinate and color space,
\be \label{eq:hopping_matrix}
H(\vec{x},\vec{y};U(t))=\sum_{j=1}^3 \left(U_j(\vec{x},t)\delta_{\vec{x}+a\hat{j},\vec{y}}+U^\dag_j(\vec{x}-a\hat{j},t)\delta_{\vec{x}-a\hat{j},\vec{y}} \right).
\ee

We also apply APE-smearing to the gauge fields that enter the hopping matrix. For the parameters $\alpha_{\rm G}$ and $n_{\rm G}$ of the Gaussian smearing we  use the values $\alpha_{\rm G} = 4.0$ and $n_{\rm G} = 50$, optimized such as to yield a proton root mean square radius of about 0.5~fm. The APE-smearing parameters are $N_{\rm APE} = 50$ and $\alpha_{\rm APE} = 0.5$.
In our calculations we choose the source-positions $(\vec{x}_0,t_0)$ randomly  in order to decrease correlations among measurements.
%In addition, at $\vec{q}=\vec{0}$, the symmetries of the action and the anti-periodic boundary conditions in the temporal direction for the quark fields imply that $G_+^{\rm 2pt}(t_s) = -G_-^{\rm 2pt}(T-t_s)$, where $T$ is the temporal extent of the lattice. Therefore, in order to decrease errors we average correlators in the forward and backward direction and define $G^{\rm 2pt}(t_s) = \frac{1}{2}\left[ G_+^{\rm 2pt}(t_s) - G_-^{\rm 2pt}(T-t_s)\right]$. Moreover, we average over the proton and the neutron correlation functions, as these two particles are degenerate in the TMF lattice formulation, thus gaining a factor of $4\times$ in the total statistics.

\begin{figure}[h]
\begin{minipage}{7.5cm}
\center
\includegraphics[width=\textwidth]{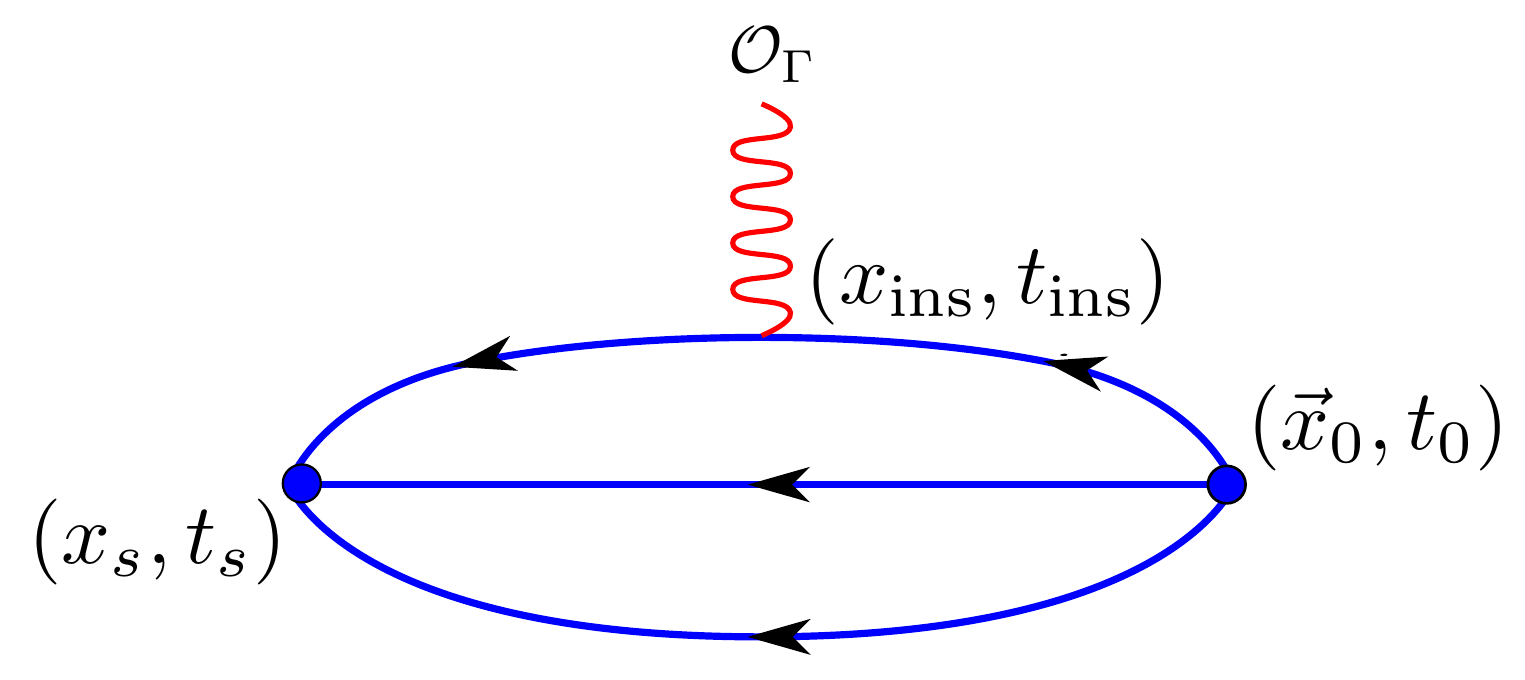}
\end{minipage}\hfill
\begin{minipage}{7.5cm}
\center
\includegraphics[width=\textwidth]{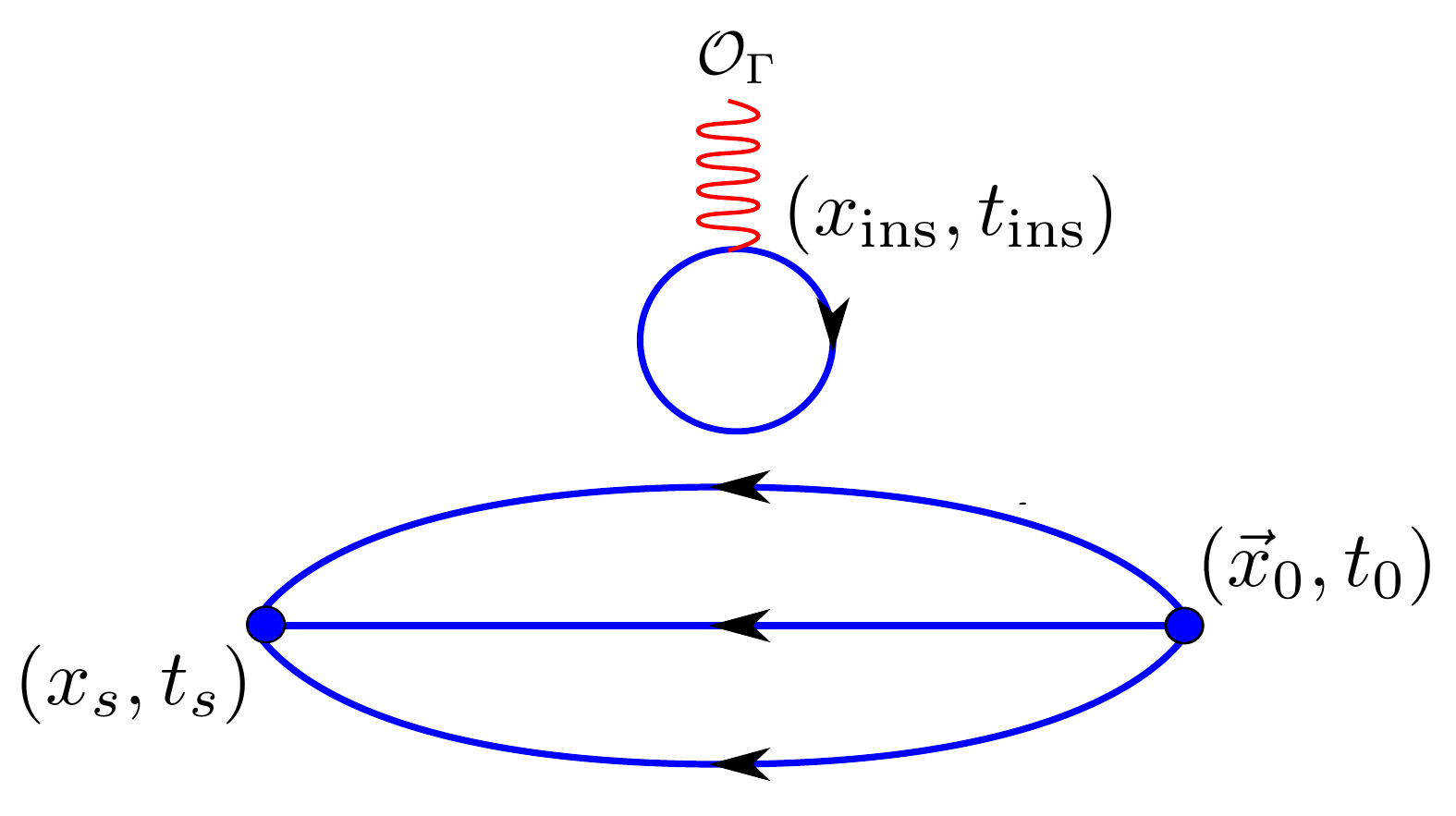}
\end{minipage}
\caption{Diagrams of a connected (left) and disconnected (right) three-point function.}
\label{fig:3pt_diagrams}
\end{figure}

As already mentioned, for  isovector quantities the disconnected contributions cancel in the isospin limit up to lattice artefacts, which we expect to be small for the twisted mass clover-improved  action used here.  In order to evaluate the connected three-point function, shown diagrammatically on the left panel of \fig{fig:3pt_diagrams}, we use the sequential inversion approach through the sink~\cite{Dolgov:2002zm}. Within this method, the sum over the sink spatial coordinates, $\vec{x}_s$, in \eq{eq:3ptfunc} is carried out through an inversion of the Dirac operator 
with an appropriately constructed source that combines the two forward propagators with the projector and the quantum numbers of the interpolating field at the sink. This so-called sequential propagator   is thus required per choice of the sink time coordinate $t_s$ and sink projector, whereas all insertion times as well as any insertion operator  can be obtained practically without additional computational cost. We perform inversions for five sink-time slices, namely $t_s/a = 10,12,14,16$ and $18$, which correspond to $t\approx0.9-1.7$~fm in physical units, for the scalar operator where excited states contributions were found to be  significantly large and three sink-times for the tensor operator, $t_s/a = 10,12$ and $14$. We use four separate projectors, namely $P^4$ and $P_k, \,k=1,2,3$ as given in Eqs.~(\ref{eq:proj_4}) and~(\ref{eq:proj_pol}).

For the isoscalar as well as the purely disconnected strange and charm quantities one needs to compute the disconnected quark loop and appropriately combine it with the two-point function in order to construct the disconnected three-point function, depicted on the right panel of~\fig{fig:3pt_diagrams}. The disconnected quark loop for a general $\gamma$-structure $\Gamma$ is of the form
\be\label{eq:disc_loop}
\mathcal{L}^{(f)}(\Gamma,t) = \sum_{x_{\rm ins}} \tr \left[G_f(x_{\rm ins};x_{\rm ins})\Gamma\right]\;,
\ee
where $G_f(x;y)$ is the propagator of the quark flavor $f$. The sum over all the spatial coordinates $x_{\rm ins}$ implies that one needs to evaluate the so-called all-to-all propagator. It is apparent that it is prohibitively expensive to calculate all-to-all propagators in an exact manner, as this would require $L^3$ inversions of the Dirac matrix per quark flavor, compared to two inversions per quark flavor required for the connected three-point function. A feasible alternative is to use stochastic techniques~\cite{Bitar:1989dn} in order to obtain an unbiased estimate of $G_f(x_{\rm ins};x_{\rm ins})$, at the cost of introducing stochastic error in the calculation. Briefly, this is usually done by generating a set of $N_r$ stochastic sources $|\xi_r\>$ randomly filled with $\mathbb{Z}_4$-noise. Then one solves $M|s_r\> = |\xi_r\>$ for $|s_r\>$ and calculates
\be\label{eq:stoch_prop}
G \equiv M_E^{-1} = \frac{1}{N_r}\sum_{r=1}^{N_r}|s_r\> \<\xi_r| \approx M^{-1}\;.
\ee
For $N_r\rightarrow \infty$, \eq{eq:stoch_prop} provides an unbiased estimate of the all-to-all propagator. The number $N_r$ required in order to sufficiently suppress the stochastic noise depends on the observable, but in general $N_r\sim \mathcal{O}(10^3)$, which is much smaller than $L^3$, hence this calculation is computationally attainable.

As already mentioned, TMFs have a  property that allows to reduce the gauge noise of disconnected quark loops. At first we remark that the isoscalar combination of a flavor doublet of the scalar operator transforms into an isovector of the pseudo-scalar operator in the twisted basis, i.e. $\bar{u}u+\bar{d}d = i(\bar{\phi}_u\gamma_5\phi_u - \bar{\phi}_d\gamma_5\phi_d)$, where $u,d$ are the quark fields in the physical basis and $\phi_u,\phi_d$ are the quark fields in the twisted basis. The disconnected quark loop for the scalar operator in the twisted basis is then given by
\be\label{eq:loop_tb}
\mathcal{L}^{(u+d)}(\mathbb{1},t) = \sum_{x_{\rm ins}} \tr \left[G_u(x_{\rm ins};x_{\rm ins})+ G_d(x_{\rm ins};x_{\rm ins})\right] \longrightarrow \sum_{x_{\rm ins}} \tr \left[i\gamma_5\left(G_{\phi_u}(x_{\rm ins};x_{\rm ins})- G_{\phi_d}(x_{\rm ins};x_{\rm ins})\right)\right]\;,
\ee
which, when utilizing the TMF properties becomes
\be\label{eq:loop_2}
\mathcal{L}^{(u+d)}(\mathbb{1},t) = 2\mu_l \sum_{x_{\rm ins},y} \tr \left[G_{\phi_d}(x_{\rm ins};y)G_{\phi_d}^\dag(y;x_{\rm ins})\right]\;.
\ee
From this transformation, known as the one-end trick~\cite{Michael:2007vn,Foster:1998vw,McNeile:2006bz}, two main advantages emerge. The first is that the gauge fluctuations are significantly reduced due to the $\mu_l$ factor, which is $\mathcal{O}(10^{-3})$. The second one is that the sum of $V$ terms (sum over $y$ in~\eq{eq:loop_2}) that appears in transforming the subtraction of propagators into a multiplication, which increases the signal-to-noise ratio from $1/\sqrt{V}$ to $V/\sqrt{V^2}$. The two benefits emerging from the one-end trick yield a large reduction in the errors for the same computational cost.

The final expression for the disconnected quark loop of the isoscalar combination of the scalar operator is obtained by using the property
\be\label{eq:stoch_property}
\frac{1}{N_r} \sum_r |\xi_r\>\<\xi_r| = \mathbb{1} + \mathcal{O}\left(\frac{1}{\sqrt{N_r}}\right)
\ee
of the noise sources. Inserting this into~\eq{eq:loop_2}, and noting that $|s_r\> = G|\xi_r\>$ we get
\be\label{eq:loop_final}
\mathcal{L}^{(u+d)}(\mathbb{1},t) = 2\mu_l \frac{1}{N_r} \sum_{r=1}^{N_r} \<s_r|s_r\> + \mathcal{O}\left(\frac{1}{\sqrt{N_r}}\right)\;,
\ee
where with the bra-ket notation, a trace over spatial volume, spin and color indices must be realized. Similarly, the isoscalar tensor operator in the twisted basis transforms as $\sigma_{\mu\nu}\mathbb{1}\rightarrow i\gamma_5\sigma_{\mu\nu}\tau^3$, where $\mathbb{1}$ and $\tau^3$ act in flavor space. Following the same procedure for the tensor operator we obtain the expression
\be\label{eq:loop_final_tensor}
\mathcal{L}^{(u+d)}(\sigma_{\mu\nu},t) = 2\mu_l \frac{1}{N_r} \sum_{r=1}^{N_r} \<s_r|\sigma_{\mu\nu}|s_r\> + \mathcal{O}\left(\frac{1}{\sqrt{N_r}}\right)
\ee
with the same noise reduction benefits.

Regarding the scalar matrix element of the strange and charm quarks, we use the heavy doublets in the twisted basis to construct the pseudoscalar current $\frac{i}{2}(\bar{\phi}_{f^+}\gamma_5\phi_{f^+} - \bar{\phi}_{f^-}\gamma_5\phi_{f^-})$, where as already mentioned $f=s,c$ and $f^\pm$ refers to taking $\pm \mu_f$. Considering both $f^+$ and $f^-$ to construct these isovector-like combinations in the twisted basis, allows us to take full advantage of the one-end trick and increase the signal-to-noise ratio of the disconnected quark loops in order to obtain the loops $\mathcal{L}^{(s)}$ and $\mathcal{L}^{(c)}$. Similar procedure is followed for the tensor matrix element, as in the light quark loops. Namely, apart from a factor of $1/2$ the same expressions as in Eqs.~(\ref{eq:loop_final}) and~(\ref{eq:loop_final_tensor}) are derived for the heavy quarks, where $|s_r\>$ are obtained by inverting the twisted-mass Wilson-clover operator with the corresponding heavy quark mass.

\begin{figure}[h]
\begin{minipage}{8.5cm}
\center
\includegraphics[width=1.01\textwidth]{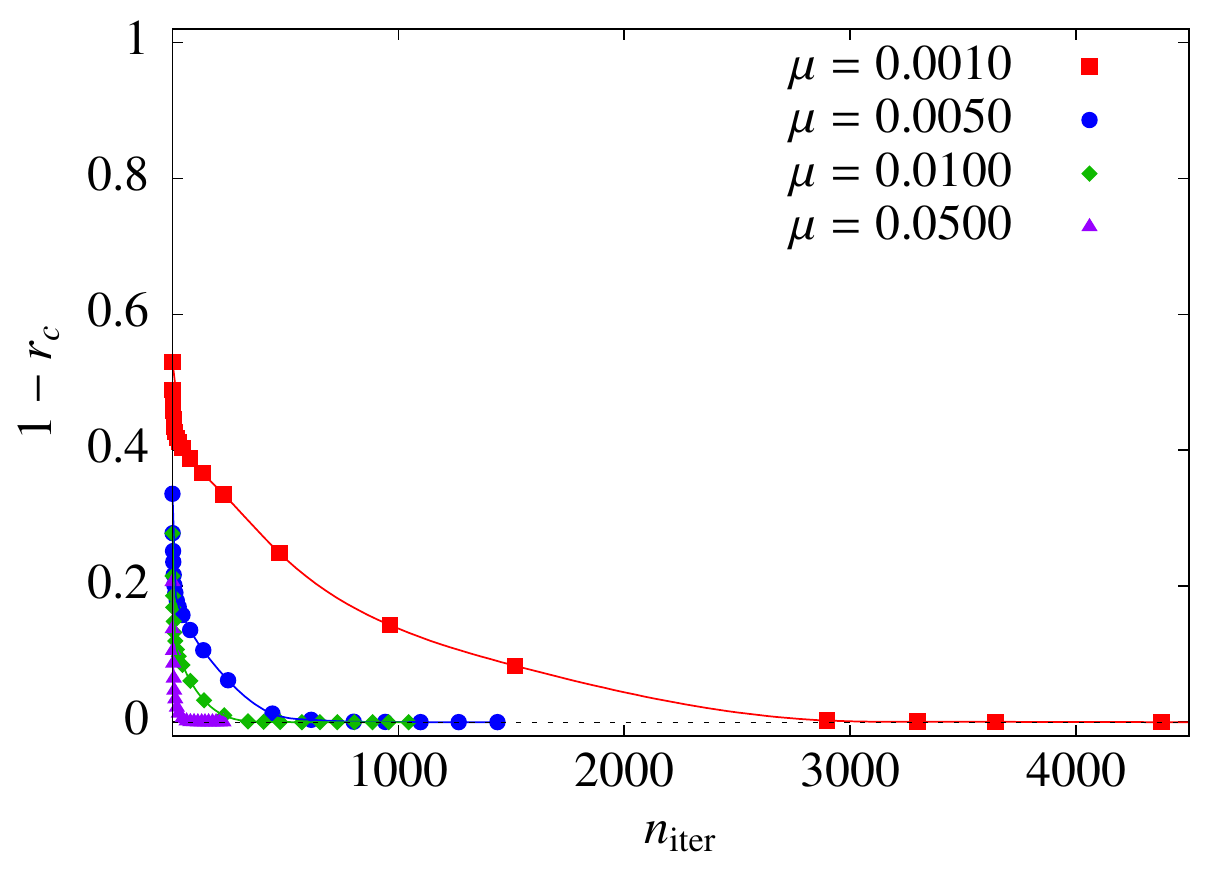}
\end{minipage}%
\hfill
\begin{minipage}{8.5cm}
\center
\includegraphics[width=\textwidth]{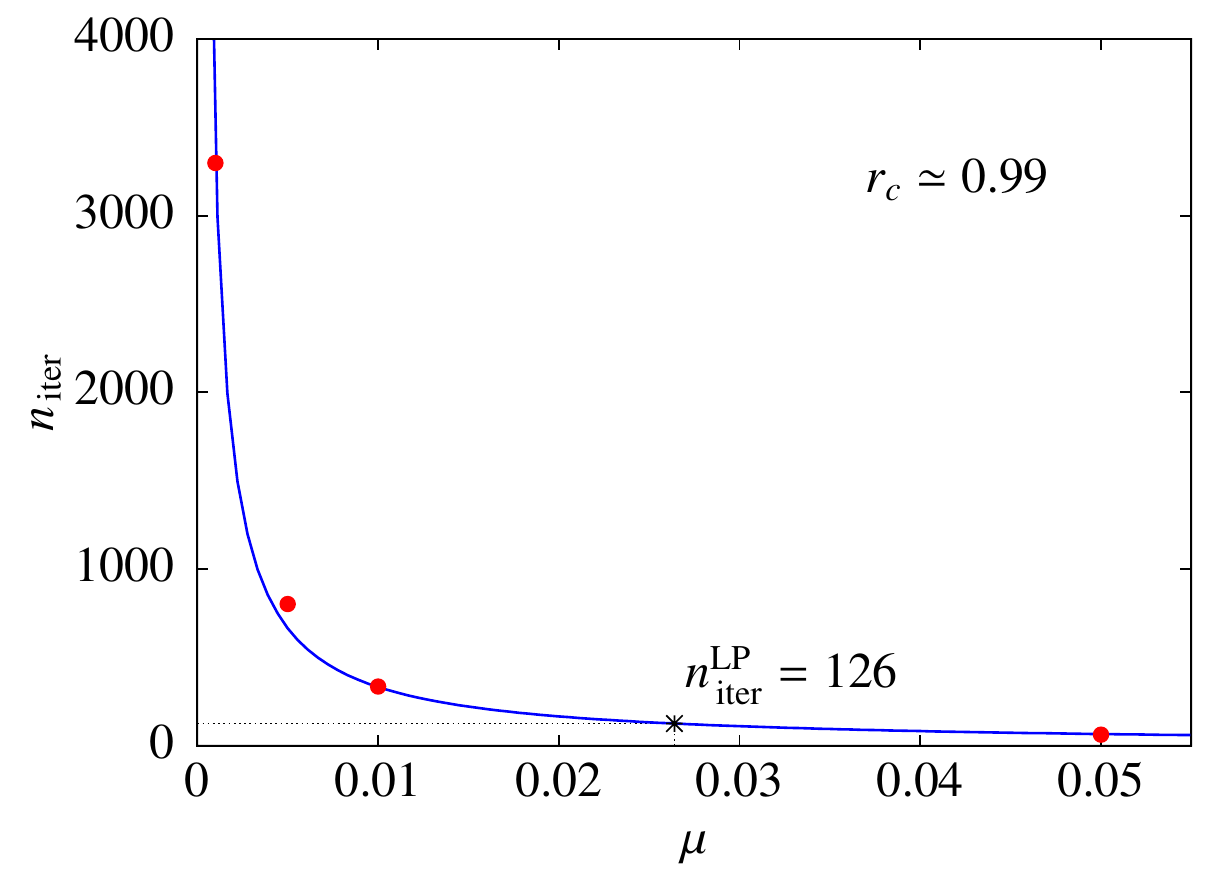}
\end{minipage}%
\hfill
\caption{Left: The correlation $r_c$ between the high precision and low precision propagators for several values of the twisted mass parameter. Right: Interpolation of our data for determining the optimal value of $n_{\rm iter}^{\rm LP}$ for the strange quark at $r_c\simeq 0.99$. A similar procedure was followed for the charm quark.}
\label{fig:TSM_tune}
\end{figure}

In addition, for the strange and charm loops we use the \emph{Truncated Solver Method} (TSM), which provides a way to increase $N_r$ at a reduced computational cost. Within this method, a large number of stochastic sources inverted to low-precision and a small number inverted to high-precision are combined to estimate the all-to-all propagator~\cite{Bali:2009hu,Alexandrou:2012zz} according to
\be\label{eq:tsm_expr}
M_{E_{\rm TSM}}^{-1} = \frac{1}{N_{\rm LP}}\sum_{j=N_{\rm HP}+1}^{N_{\rm HP}+N_{\rm LP}} |s_j\>_{\rm LP} \<\xi_j| + \frac{1}{N_{\rm HP}}\sum_{r=1}^{N_{\rm HP}} \left[|s_r\>_{\rm HP} - |s_r\>_{\rm LP} \right]\<\xi_r|\;,
\ee
where the first term is similar to~\eq{eq:stoch_prop} and the second term corrects for the bias introduced by using low-precision stochastic propagators. The parameters that need to be tuned are the exact number of low- and high-precision sources as well as the low-precision criterion, such that the all-to-all propagator remains unbiased. The latter can be set either by a relaxed stop condition for the residual of the CG algorithm, for instance $|\hat{r}|<10^{-2}$, or equivalently, by fixing the number of iterations. The goal is to choose the ratio $N_{\rm LP}/N_{\rm HP}$ as large as possible, while still ensuring that the final result is unbiased and that $r_c \simeq 1$, where $r_c$ is the correlation between the $N_{\rm HP}$ propagators in low and high precision. It is customary to set $N_{\rm LP}$ as the number of sources that would be used if the standard stochastic method was to be employed instead of TSM, and then increase the number of $N_{\rm HP}$ until the bias is corrected, see e.g. Refs.~\cite{Alexandrou:2012zz,Abdel-Rehim:2013wlz,Alexandrou:2013wca}. We set the low-precision criterion such that $r_c\simeq 0.99$, which is sufficiently large for the purposes of our calculation. We show the dependence of $r_c$ as a function of the CG iterations on the left panel of~\fig{fig:TSM_tune} for various values of the twisted mass parameter $\mu$. To determine the exact number of iterations for each $\mu$-value, we interpolate our data as demonstrated on the right panel of~\fig{fig:TSM_tune} for the case of the strange quark. From this procedure we find $n_{\rm iter}^{\rm LP} = 126$ for the strange quark. Following a similar procedure for the charm quark we find $n_{\rm iter}^{\rm LP} = 9$. However, by fixing the number of iterations the exact residual might differ amongst the stochastic sources inverted. In order to avoid that, we equivalently opt to set $|\hat{r}_{\rm LP}| = 10^{-3}$ as the low-precision criterion in both cases, which still satisfies the condition $r_c\simeq 0.99$ and yields iteration numbers very close to the values obtained from the TSM tuning procedure. The values of $N_{\rm HP}$ and $N_{\rm LP}$ for the strange and charm quarks are listed in~\tbl{Table:statistics}. We remark here that applying the TSM method for the light quarks is not as beneficial since, as one can see from the left plot of~\fig{fig:TSM_tune}, the number of iterations required to achieve a high correlation is much larger than for the heavy quarks. In fact, in an attempt to tune the TSM parameters following the discussion of Ref.~\cite{Bali:2009hu} we found that the resulting optimal values reported minimal benefits.
%
% TSM plots

After calculating the two- and three-point functions, we then form the ratio
\be\label{eq:ratio_3pt2pt}
R_\Gamma(P,t_s,t_{\rm ins}) = \frac{G_\Gamma^{\rm 3pt}(P,\vec{0},\vec{0},t_s,t_{\rm ins})}{G^{\rm 2pt}(\vec{0},t_s)}\;.
\ee 
 In the large time limit, the unknown overlaps of the nucleon interpolating field with the nucleon ground state cancel and the ratio becomes time independent, thus the desired matrix element can be extracted from a fit to a constant. This can be realized by writing~\eq{eq:ratio_3pt2pt} on the hadron level
\be\label{eq:ratio_hadrl}
R_\Gamma(P,t_s,t_{\rm ins}) \propto \frac{\sum_{n,n'}\<J|n'\>\<n|\bar{J}\>\<n'|\mathcal{O}_\Gamma|n\> e^{-E_{n'}(t_s-t_{\rm ins})}e^{-E_n(t_{\rm ins}-t_0)}}{\sum_n |\<J|n\>|^2 e^{-E_n(t_s-t_0)}}\;,
\ee
where $|n\>$ is the $n^{\rm th}$ eigenstate of the QCD Hamiltonian with the quantum numbers of the nucleon, and $E_n$ is the rest frame energy of that state. We note that when $|n\>=|N\>$ and $|n'\>=|N\>$, where $|N\>$ is the nucleon ground state, the desired matrix element $\<N|\mathcal{O}_\Gamma|N\>$ appears in~\eq{eq:ratio_hadrl}. The exponential terms containing energies of excited states become small compared to the matrix element and compared to unity when $t_{\rm ins}-t_0 \gg 1$ and $t_s-t_0 \gg 1$, in which case the ratio reduces to the desired ground state matrix element.

\subsection{Excited states investigation}

In order to assure that the extracted matrix element corresponds to the nucleon ground state, we employ three methods to assess whether contributions due to excited states to the ratio of~\eq{eq:ratio_hadrl} are sufficiently suppressed.

\begin{figure}[h]
\begin{minipage}{8.5cm}
\center
\includegraphics[width=\textwidth]{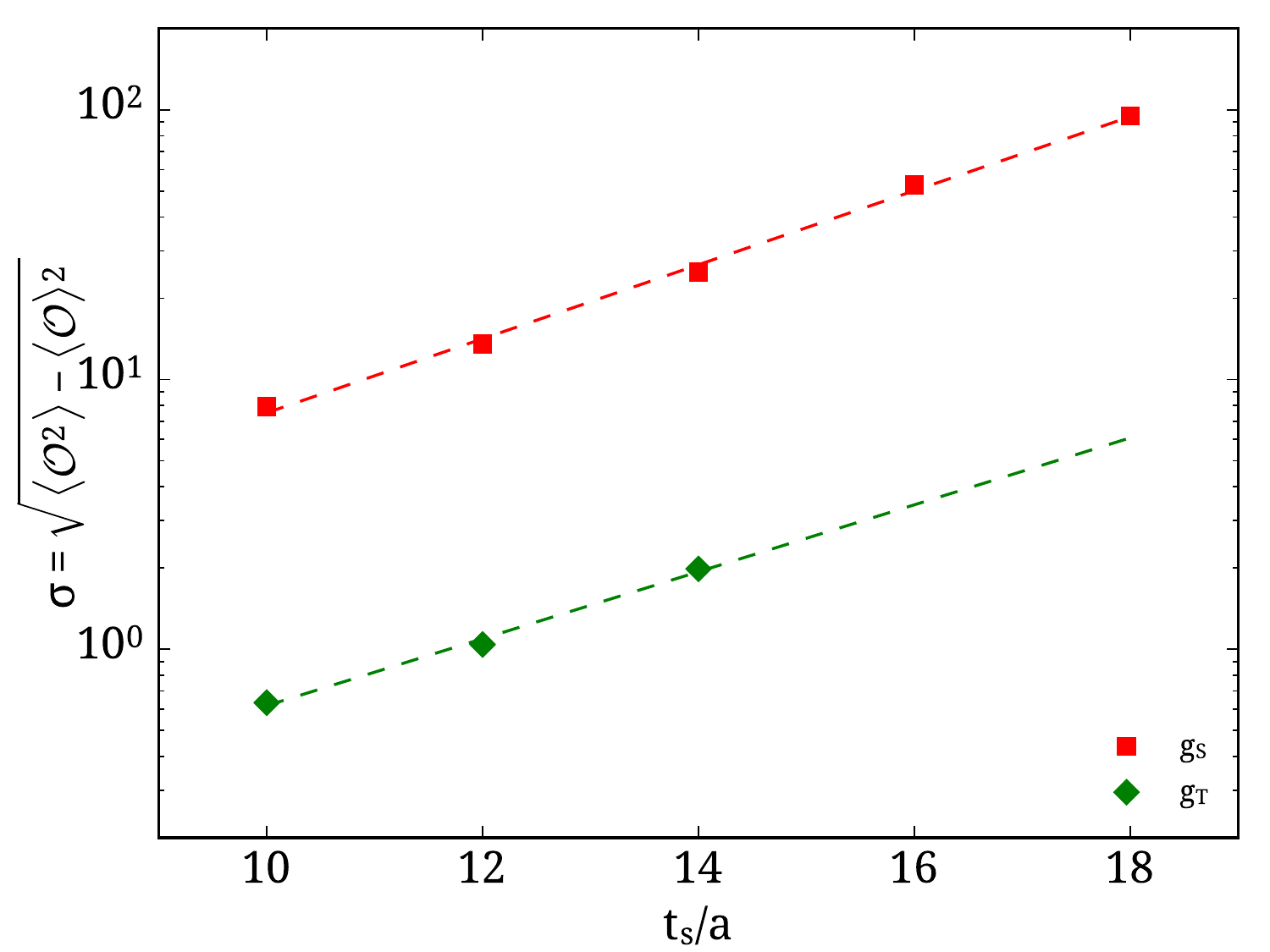}
\end{minipage}\hfill
\center
\caption{The variance as a function of the sink-source time separation for the isovector scalar and the tensor charges.}
\label{fig:var}
\end{figure}

The first method, known as \emph{the plateau method}, is  commonly used  in extracting matrix elements. One computes the three-point function for several sink-source time separations and examines the time dependence of  the ratio given by~\eq{eq:ratio_hadrl}. If $\Delta(t_{\rm ins}-t_0) \gg 1$ and $\Delta (t_s-t_0) \gg 1$, where $\Delta = E_1-E_0$ then contributions from excited states are expected to be small and ground dominance leads to a time-independent region (plateau). However, due to the fact that the approach to the plateau value is not monotonic, identifying the plateau region can become a difficult task. For this reason, several $t_s$ need to be computed for each matrix element, which should demonstrate a convergence towards a single constant  plateau value as $t_s$  is increased. As already mentioned, within the sequential inversion through the sink, the sink time-slice is fixed, thus a new set of inversions is required for each new value of $t_s$. An additional issue arising as the sink-source time separation increases is that the error to signal increases exponentially, as shown in Fig.~\ref{fig:var}. Therefore,  the statistics required for  constant error increase exponentially making the calculation computationally very demanding. 

In the plateau region one fits the ratio
\be\label{eq:ratio_plat}
R_\Gamma(P,t_s,t_{\rm ins}) \xrightarrow[\Delta(t_s-t_0) \gg 1]{\Delta(t_{\rm ins}-t_0)\gg1} \Pi_\Gamma(P)
\ee
over $t_{\rm ins}$ to a constant  to obtain the desired matrix element. The procedure is repeated for several increasing values of $t_s$ until the plateau value does not change, in order to ensure that the contaminations from excited states are suppressed.
The scalar and tensor charges of the nucleon ground state, at zero momentum transfer in the large Euclidean time, are then extracted from the corresponding ratios
\bea
\Pi_S(P^4) &=& \frac{g_S}{2} \nonumber\\
\Pi_T^{ij}(P_k) &=& \epsilon^{ijk}\frac{g_T}{2}.
\label{eq:plat_gST}
\eea

The second technique we employ is the \emph{summation method}~\cite{Maiani:1987by}, which has been applied in a number of recent calculations concerning nucleon charges~\cite{Capitani:2012gj,Abdel-Rehim:2015owa,Abdel-Rehim:2016won}. In this approach, a sum of the ratio over the insertion time $t_{\rm ins}$ is performed
\be\label{eq:summ_meth}
R_\Gamma^{\rm sum} (P,t_s) = \sum_{t_{\rm ins}=t_0+a}^{t_s-a} R_\Gamma(P,t_s,t_{\rm ins})\;.
\ee
From~\eq{eq:ratio_hadrl} and keeping terms up to $E_1$ one can see that the sum over the exponentials is a geometric series, thus it can be easily carried out and reads
\be\label{eq:summ_meth_2}
R_\Gamma^{\rm sum} (P,t_s) =C + (t_s-t_0)\mathcal{M} + \mathcal{O}(e^{-\Delta(t_s-t_0)})+\cdots\;,
\ee
where $C$ is a constant independent of $t_s$ and $\mathcal{M} \equiv \<N|\mathcal{O}_\Gamma|N\>$ is the desired matrix element. The matrix element $\mathcal{M}$ is then the slope of a straight line fit of $R_\Gamma^{\rm sum} (P,t_s)$ w.r.t. $(t_s-t_0)$. In general, since we now fit over two parameters $C$ and $\mathcal{M}$, the summation method has larger errors on the matrix element. 

The third approach to extract the desired matrix element is the so-called \emph{two- or three-state fit}. Within this method the contributions due to the first or second excited state are taken into account. In this analysis we consider terms involving the first excited state i.e. we perform a two-state fit.
We take into account several values of the sink-source separation and perform a simultaneous combined fit with respect to $t_{\rm ins}$ and $t_s$, taking into account all terms that involve the ground state and the first excited state. From~\eq{eq:ratio_hadrl}, one can see that considering all terms up to the first excited state gives the following expression for the three-point function
\bea\label{eq:3pt_2state}
G_\Gamma^{\rm 3pt} &=& A_{00}e^{-E_0(t_s-t_{0})} + A_{01} e^{-E_0(t_s-t_{\rm ins})}e^{-E_1(t_{\rm ins}-t_0)}  \nonumber\\
&+& A_{10}e^{-E_1(t_s-t_{\rm ins})}e^{-E_0(t_{\rm ins}-t_0)} + A_{11}e^{-E_1(t_s-t_{0})}\;,
\eea
where $E_0=m_N$, the mass of the nucleon, $A_{nm} = \<J|n\>\<m|\bar{J}\> \< n|\mathcal{O}_\Gamma|m\>\;, A_{nm}\in \mathbb{R}$, and we note that $A_{01}=A_{10}$. Similarly, the expression for the two-point function is given by
\be\label{eq:2pt_2state}
G^{\rm 2pt} = c_0 e^{-E_0(t_s-t_0)} + c_1 e^{-E_1(t_s-t_0)}\;,
\ee
where $c_n = |\<J|n\>|^2$. In the above expressions, $|0\>$ denotes the ground state of the nucleon and $|1\>$ the first excited state. We perform a simultaneous  fit to the three- and two-point functions given in  Eqs.~(\ref{eq:3pt_2state}) and~(\ref{eq:2pt_2state}) that includes seven fit parameters, namely $A_{00}, A_{01}, A_{11}, c_0, c_1, E_0$ and $E_1$. The desired matrix element $\mathcal{M}$ is then obtained from
\be\label{2state_matel}
\mathcal{M}\equiv \<0|\mathcal{O}_\Gamma|0\> = \frac{A_{00}}{c_0}\;.
\ee
These fits  are very robust and enable us to extract the excited state contribution accurately. We note that for a consistency we also  perform a direct fit to the ratio of~\eq{eq:ratio_hadrl}, which includes five fit parameters instead of seven. We find consistent results of the matrix element between the two fits, albeit the errors from the latter fit are larger. We thus use the results extracted from the seven-parameter fit.

%The overlaps of the nucleon interpolating field with the ground state and the first excited state are treated as fit parameters, one of which also includes $\mathcal{M}$. After the fit, an appropriate expression is formed which cancels these overlaps in order to isolate the matrix element $\mathcal{M}$.

We consider that excited states are sufficiently suppressed when the three methods mentioned above yield consistent values for $\mathcal{M}$, and take the plateau fit at the $t_s$ for which this agreement holds as our final value. We give as a systematic error due to excited state effects the difference between the mean values extracted from the plateau and two-state fit. 

\subsection{Finite lattice spacing and volume effects}

Since we are using a single ensemble we cannot directly evaluate lattice artifacts due to the finite lattice spacing and volume. However, we have computed
similar matrix elements using  $N_f=2$~\cite{Alexandrou:2010hf,Alexandrou:2011db,Alexandrou:2011nr} and $N_f=2+1+1$~\cite{Alexandrou:2013joa} ensembles for higher than physical pion masses,  where we compared results using three different lattice spacings and different volumes.
Within our statistical errors, results on for example  the axial charge
were found to be consistent for lattice spacings between $a\sim 0.9$~fm and $a\sim 0.6$~fm. A continuum extrapolation yielded a value consistent with
that determined with the ensemble at $a\sim 0.9$~fm~\cite{Alexandrou:2010hf}. We thus expect cut-off effects to be small for
our improved action.
Similarly, comparing the axial charge extracted for $  Lm_\pi = 3.3$ and $Lm_\pi=4.3$ we observed no detectable volume effects. A further indication
that volume effects
are under control is based on 
a study of the nucleon $\sigma$-terms~\cite{Abdel-Rehim:2016won} extracted from the same scalar matrix element using the same ensemble employed here.
Our value for the $\sigma_{\pi N}$ is in agreement with the result of Ref.~\cite{Durr:2015dna} obtained from an extensive analysis using the Feynman-Hellmann theorem which was corrected for finite volume effects. 
This leads us to  expect that finite volume effects are reasonably small, although an investigation of volume effects at high accuracy is called for at
the physical point. We are
currently investigating volume effects on these quantities  using a spatial lattice size of $L_s=64$
at the same pion mass as the one of this work.
%==========================================================================
%==========================================================================

\section{Results}
\label{sec:results}

In this section we present our results for the light, strange and charm scalar and tensor charges. 
Connected contributions are computed  for five source-sink separations for the unpolarized projector (\eq{eq:proj_4}) used for the scalar charges, namely $t_s/a = 10,12,14,16$ and $18$, which correspond to about $0.9,1.1,1.3,1.5$ and $1.7$~fm. For the polarized projector (\eq{eq:proj_pol}) applied for the tensor charges we use $t_s/a = 10,12$ and $14$, which as becomes clear in the following, prove sufficient for our analysis.
%where we have set $t_0=0$.
The disconnected quark loops are calculated for all time slices. In~\tbl{Table:statistics} we summarize the statistics of our calculation for both the connected and disconnected contributions. Also listed is the number of high- and low-precision stochastic sources used in the TSM for the strange and charm charges. The statistics for $t_s/a = 10,12$ and $14$ analyzed in this study are about six times more than Ref.~\cite{Abdel-Rehim:2015owa}. The two larger time separations, namely $t_s/a = 16$ and $18$ that we have introduced in this work serve as a further check for excited state effects.
%Furthermore, we include the disconnected contributions to the isoscalar quantities and compute the strange and charm nucleon charges. 
%
\begin{table}
\begin{center}
\renewcommand{\arraystretch}{1.2}
\renewcommand{\tabcolsep}{3.5pt}
\begin{tabular}{cccc||cccccc}
\multicolumn{4}{c||}{Connected: three-point} & \multicolumn{6}{c}{Disconnected} \\
\hline\hline
$t_f/a$ & $N_{\rm conf}$ & $N_{\rm src}$ & $N_{\rm tot}$ & Flavor & $N_{\rm src}$ & $N_{\rm conf}$ & $N_{\rm tot}$ & $N_{\rm HP}$ & $N_{\rm LP}$ \\
\hline
10,12,14 & 579 & 16 & 9264  & light   & 100 & 2137 & 213700 & 2250 &  -  \\
16       & 542 & 88 & 47696 & strange & 100 & 2153 & 215300 &   63 & 1024  \\
18       & 793 & 88 & 69784 & charm   & 100 & 2153 & 215300 &    5 & 1250  \\
\hline
\end{tabular}
\caption{The statistics of our calculation. $N_{\rm conf}$ is the number of gauge configurations analyzed and $N_{\rm src}$ is the number of source positions per configuration. With $N_{\rm tot}$ we denote the total number of statistics, i.e. $N_{\rm tot} = N_{\rm conf}\times N_{\rm src}$. In the case of the disconnected contributions, $N_{\rm src}$ refers to the number of two-point functions. Also given here is the number of high-precision stochastic vectors, $N_{\rm HP}$, produced for the loops as well as the number of low-precision vectors produced, $N_{\rm LP}$, in the cases that the TSM is employed.}
\label{Table:statistics}
\end{center}
\end{table}

In order to renormalize the scalar and tensor matrix elements it is sufficient to evaluate the renormalization functions  of the  non-singlet and singlet scalar and tensor quark bilinears, which for maximally twisted fermions are given by  $Z_P$ and $Z_T$, for the gauge ensemble we use in this work. We determine them non-perturbatively in the $\overline{\rm MS}$-scheme at a scale of 2~GeV. 
Details on the computation of the non-singlet renormalization functions are given in Ref.~\cite{Alexandrou:2015sea}.
For flavor singlet operators, disconnected fermion lines lead to significant increase in the computational effort. In order to calculate the renormalization coefficients non-perturbatively, we consider the bare vertex functions~\cite{Gockeler:1998ye}
\begin{equation}
G^{ns}(p) = \frac{a^{12}}{V} \sum_{x,y,z} \langle u(x) \bar{u}(z) \Gamma d(z) \bar{d}(y) \rangle, \;\; G^{s}(p) = \frac{a^{12}}{V} \sum_{x,y,z} \langle u(x) \bar{u}(z) \Gamma u(z) \bar{u}(y) \rangle  
\end{equation}  
where $G^{ns}$ and $G^{s}$ are the non-singlet and singlet cases, respectively, and $V$ is the lattice volume. The amputated vertex function can be derived from the vertex function as
\begin{equation}
\Lambda_\Gamma(p) = (S(p))^{-1} G_\Gamma(p) (S(p))^{-1}
\end{equation}
where $S(p)$ is the propagator in momentum-space. For the singlet vertex function the disconnected contribution is amputated using one inverse propagator because the closed quark loop does not have an open leg. More details will be given in a fore-coming publication~\cite{renormNf2_paper}.
The values used to renormalize the lattice scalar and tensor matrix
elements are given in~\tbl{Table:Zfunc} for both isovector and
isoscalar quantities.
Perturbatively, the difference between the singlet and non-singlet
renormalization functions for both $Z_P$ and $Z_T$ is zero up to two loops,
as presented
in Ref.~\cite{Constantinou:2016ieh}. A non-zero difference is present
 only for the scalar and axial operators. In particular, for
the scalar operator
which breaks chirality similarly to the pseudoscalar and tensor ones,
the difference is extremely small for the Iwasaki gauge action
combined with
the value of $c_{SW}$ used in our simulations. This behavior partly
explains the small difference we find
non-perturbatively~\cite{renormNf2_paper}
as presented in~\tbl{Table:Zfunc}.
 
%(see Fig.~4 of Ref.~\cite{Constantinou:2016ie}) 
\begin{table}
\begin{center}
\renewcommand{\arraystretch}{1.2}
\renewcommand{\tabcolsep}{3.5pt}
\begin{tabular}{c|c|c}
   & $Z_P^{\overline{\rm MS}}$  &  $Z_T^{\overline{\rm MS}}$  \\
\hline\hline
Singlet      & 0.4997(38)(177)  & 0.8515(3)(51) \\
Non-singlet  & 0.5012(75)(258)  & 0.8551(2)(15) \\
\hline
\end{tabular}
\caption{Renormalization functions $Z_P$ and $Z_T$ for the gauge ensemble analyzed in this work, given in the twisted basis. $Z_P$ renormalizes the scalar operator in the physical basis, whereas $Z_T$ is the same in both bases. The first error is statistical and the second error is the systematic due to continuum extrapolation.}
\label{Table:Zfunc}
\end{center}
\end{table}

We illustrate  our results in what follows using a common format, namely we show two plots per observable: 
%At first, we note that we measure all times relative to $t_0$, i.e. $t_{\rm ins}$ now measures the source-insertion separation and $t_s$ the source-sink separation.
 In the first plot we display for each $t_s$ the ratio of~\eq{eq:ratio_3pt2pt} as a function of $t_{\rm ins}-t_s/2$, such that the midpoint time of the ratio coincides for all source-sink separations at $t_{\rm ins}-t_s/2 = 0$. We also include in the same plot the horizontal bands resulting from fitting the ratio, the summation method and the two-state fit. In the second plot we summarize the three methods we employ in the calculations, by demonstrating the convergence of our results extracted from fitting the ratio in the plateau region  as a function of $t_s$, as well as, by including the results of the summation method and the two-state fit as we vary the lowest value of $t_s$ considered in the fits, denoted by $t_s^{\rm low}$. Throughout, all errors including the error bands in the fits are calculated using jackknife resampling.

\subsection{Nucleon scalar charge}
\begin{figure}[h]
\begin{minipage}{8.5cm}
\center
\includegraphics[width=\textwidth]{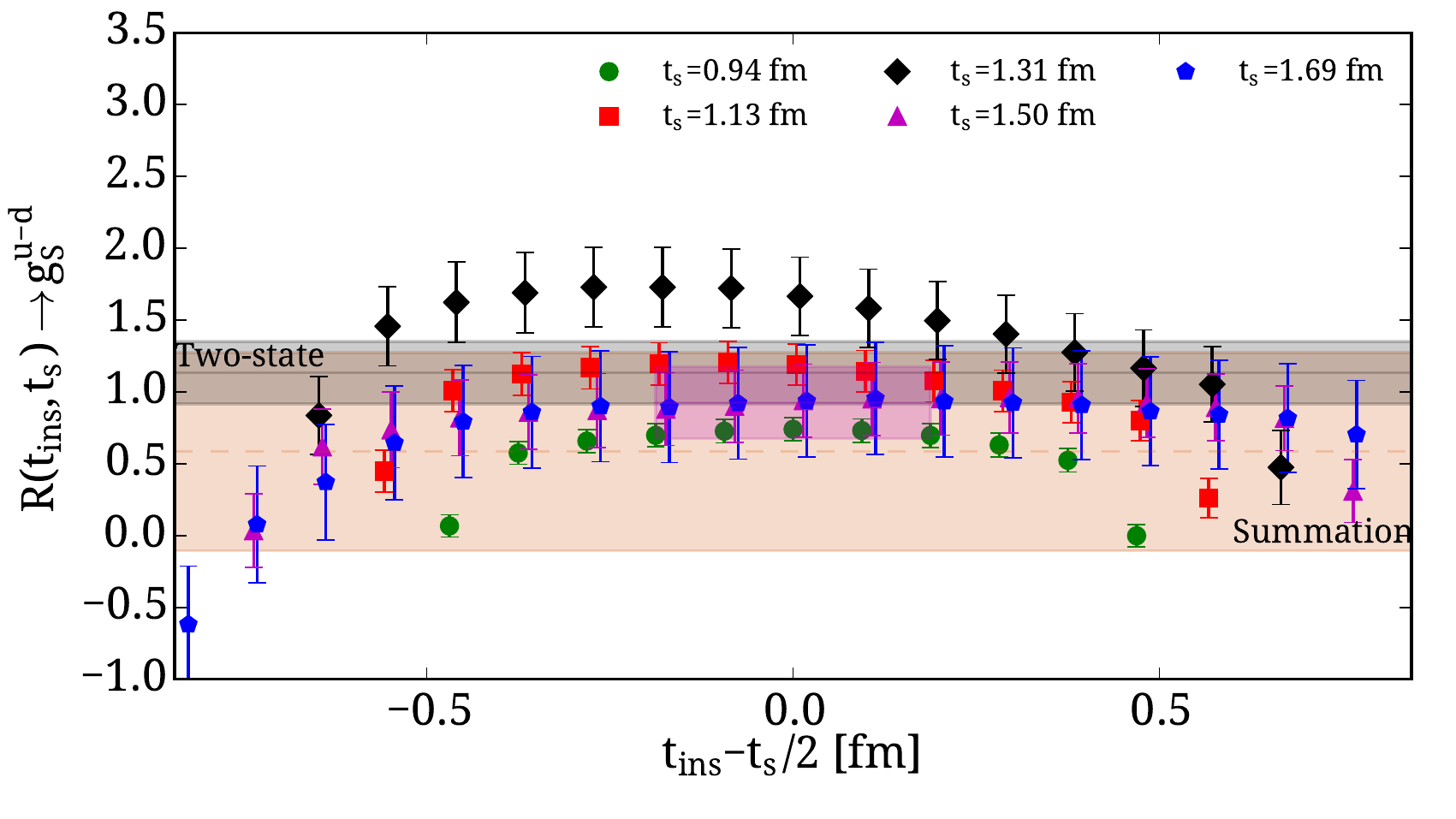}
\end{minipage}\hfill
\begin{minipage}{8.5cm}
\center
\includegraphics[width=\textwidth]{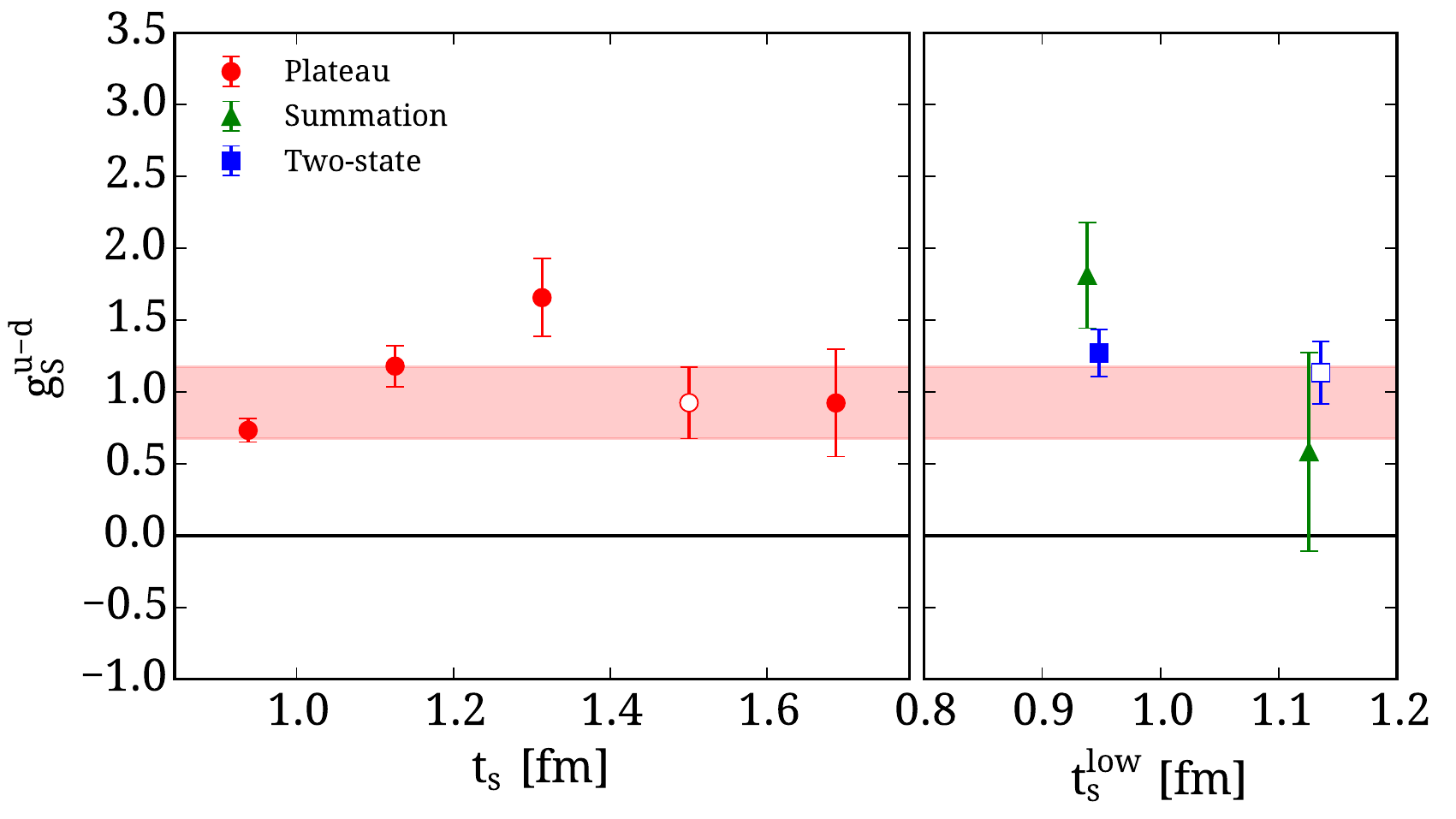}
\end{minipage}
\begin{minipage}{8.5cm}
\center
\includegraphics[width=\textwidth]{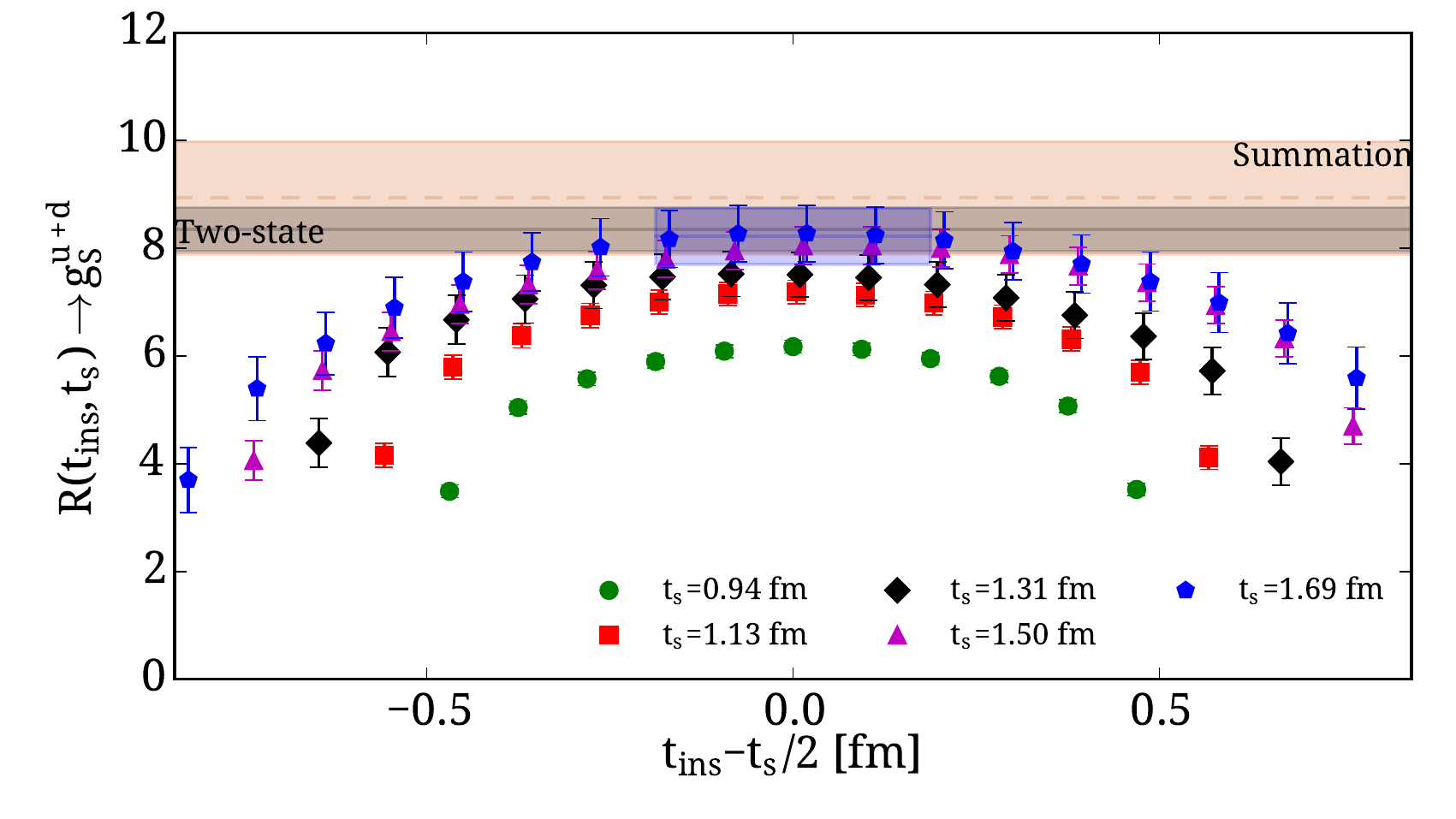}
\end{minipage}\hfill
\begin{minipage}{8.5cm}
\center
\includegraphics[width=\textwidth]{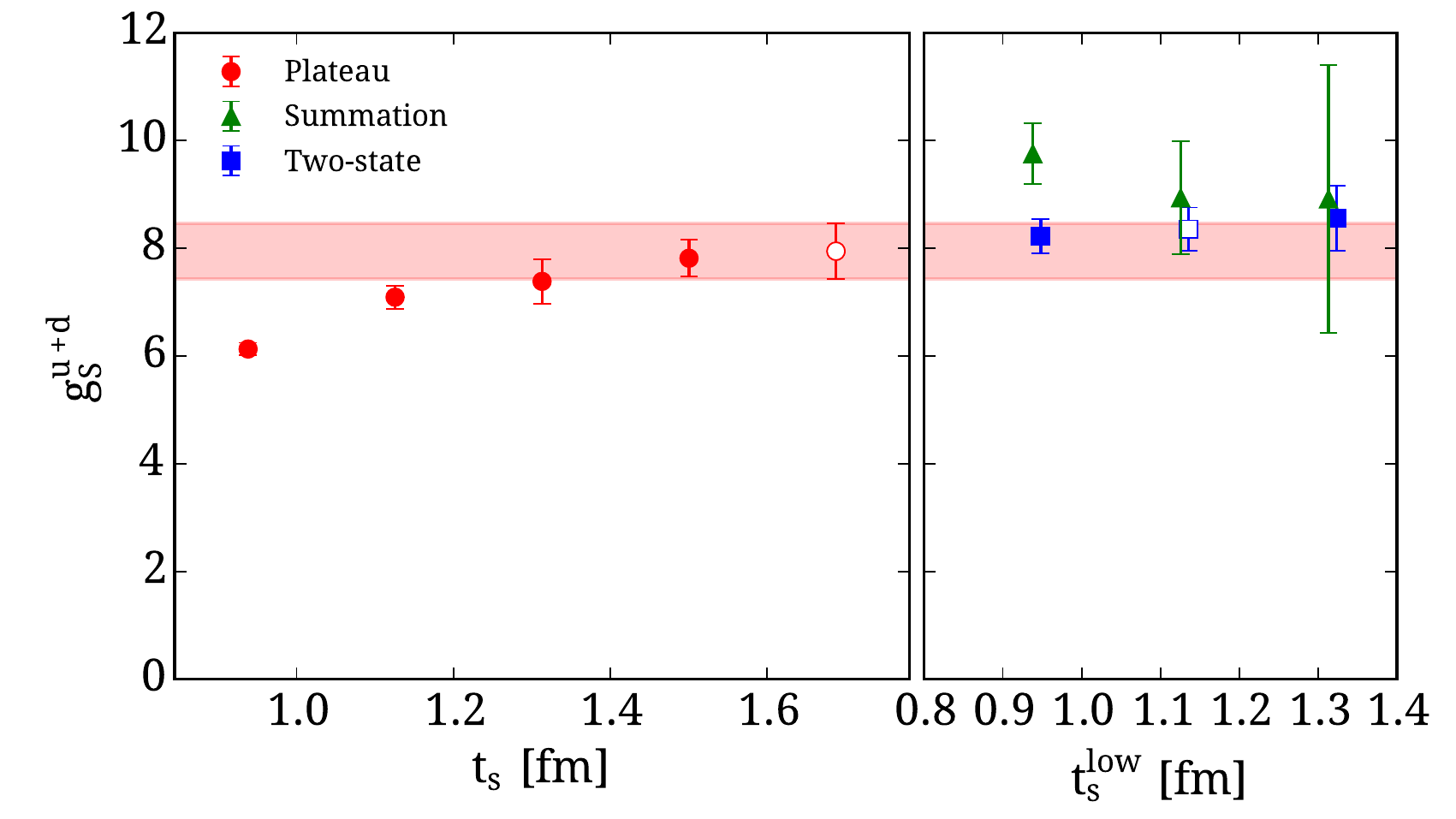}
\end{minipage}
\caption{Top left: Ratio yielding $g_S^{u-d}$ as a function of $t_{\rm ins}-t_s/2$ for source-sink separations $t_s=0.94$~fm (green circle), $t_s=1.13$~fm (red square), $t_s=1.31$~fm (black diamond), $t_s=1.50$~fm (purple triangle) and $t_s=1.69$~fm (blue pentagon). The plateau value selected is shown by the short band with the color of the corresponding $t_s$ selected, with its starting and ending points indicating the fit range used. The other two bands spanning the whole range of the plot show the results we select for the summation method (light brown) and the two-state fit (gray). Top right: Summary of our results for $g_S^{u-d}$ from the plateau fits (left column) and the summation method and two-state fit (right column). With $t_s^{\rm low}$ we denote the smallest $t_s$ in the latter two fits. The open red and blue symbols denote the selected final results from the plateau and two-state fits. The red band is the statistical error of the plateau fit. Bottom left and right: Corresponding plots for the connected contributions to $g_S^{u+d}$.}
\label{fig:gS_light_conn}
\end{figure}

In this section we discuss our results on the scalar charge of the nucleon. In \fig{fig:gS_light_conn} we show the two plots as discussed above for both the isovector scalar charge, $g_S^{u-d}$, and the connected contributions to the isoscalar scalar charge, $g_S^{u+d}$.
%As explained in the previous section, the ratio forms a plateau region when $\Delta t_{\rm ins}\gg 1$ and $\Delta (t_s-t_{\rm ins})\gg 1$. Performing a fit in the plateau region to a constant yields the plateau value from which the corresponding observable is obtained, as in Eqs.~(\ref{eq:ratio_plat}) and~(\ref{eq:plat_gST}). 
We note that the isovector scalar charge is noisy since it results from subtracting two large values. This explains the fact  that at  $t_s=1.31$~fm the plateau is consistently higher than all the rest. The statistics for this sink-source separation is the same as for the two smaller, whereas  for $t_s=1.50$~fm and $t_s=1.69$~fm  the statistics is about five and seven times more. Thus, we interpret the higher value from  the plateau as  a statistical effect. The rest of the time separations yield consistent values in the plateau region. We take the value of the plateau at $t_s=1.50$~fm, which is in agreement with the results from the summation method and two-state fit, as our final value for $g_S^{u-d}$. The difference between the value extracted from the plateau and the two-state fit is taken as a systematic error due to excited state contamination. 
 
Our previous studies of the isoscalar scalar charge, $g_S^{u+d}$,  have shown large contamination due to excited states~\cite{Abdel-Rehim:2015owa,Abdel-Rehim:2013wlz}. As can be seen from the results shown in \fig{fig:gS_light_conn}, the apparent curvature and the increasing trend in the plateau regions of the ratio as $t_s$ becomes larger confirm this observation. Both values at $t_s=1.50$~fm and  $t_s=1.69$~fm are consistent. The accuracy obtained for  $g_S^{u+d}$ allows an accurate determination at $t_s=1.69$~fm ensuring ground state dominance and it is the value we adopt. The results regarding the isovector and isoscalar scalar charges in this study corroborate the findings from Refs.~\cite{Abdel-Rehim:2015owa,Abdel-Rehim:2016won} that large source-sink separations and high statistics are required for a reasonable agreement of all three methods. We take the difference
between the plateau value at $t_s=1.69$~fm and the one 
extracted from the two-state fit starting at $t_s^{\rm low}=1.13$~fm as the systematic error due to residual excited states for both  $g_S^{u-d}$ and the connected $g_S^{u+d}$.
%
% Connected IV and IS light scalar charge

\begin{figure}[h]
\begin{minipage}{8.5cm}
\center
\includegraphics[width=\textwidth]{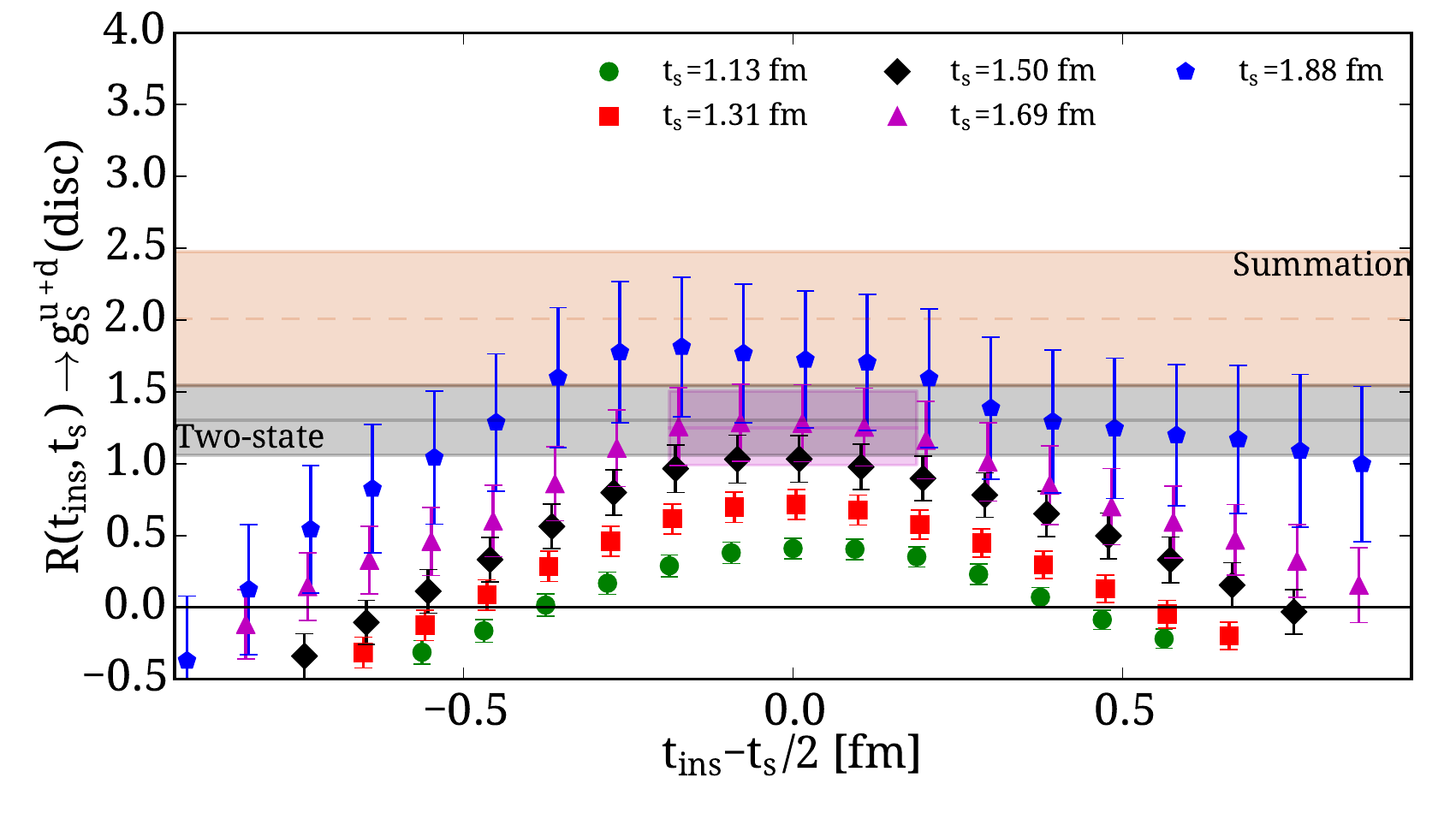}
\end{minipage}\hfill
\begin{minipage}{8.5cm}
\center
\includegraphics[width=\textwidth]{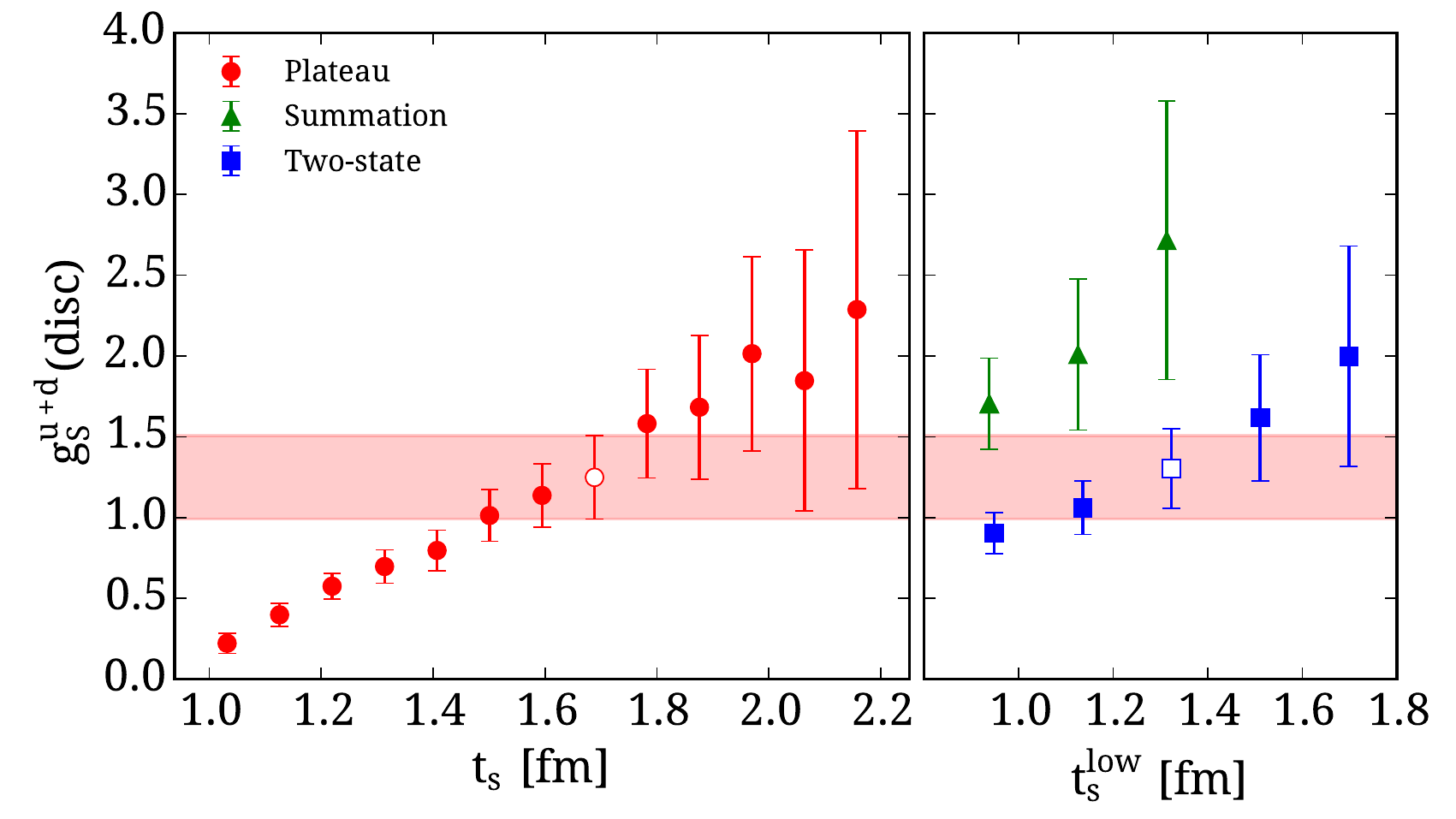}
\end{minipage}
\caption{Disconnected contributions to $g_S^{u+d}$. The notation is as in~\fig{fig:gS_light_conn}. The various values for $t_s$ shown for the plateau method are listed in the legend of the plots.}
\label{fig:gS_light_disc}
\end{figure}

We show the ratio for the disconnected contributions to $g_S^{u+d}$ in \fig{fig:gS_light_disc}. As with the connected contribution to $g_S^{u+d}$, we need to go to large time separation $t_s$ in order to suppress sufficiently the excited states, with the $t_s$-dependence being more pronounced. We note that for the disconnected quantities one can produce the ratio for all source-sink separations because the two-point function and the quark loop are produced for all time slices. We plot the dependence on $t_s$ on the right panel of~\fig{fig:gS_light_disc} where we show the plateau fits for all $t_s$ between about $1.0-2.2$~fm.  We select the plateau fit value  at $t_s=1.69$~fm as our final result for $g_S^{u+d}$, which is in agreement with the values obtained for larger values of $t_s$. This yields a contribution, which is about 15\% of the connected contribution to $g_S^{u+d}$ and is approximately the same as the percentage found for a twisted mass ensemble with a pion mass of $m_\pi=373$~MeV~\cite{Abdel-Rehim:2013wlz}. 
% Disconnected IS (light) scalar charge

\begin{figure}[h]
\begin{minipage}{8.5cm}
\center
\includegraphics[width=\textwidth]{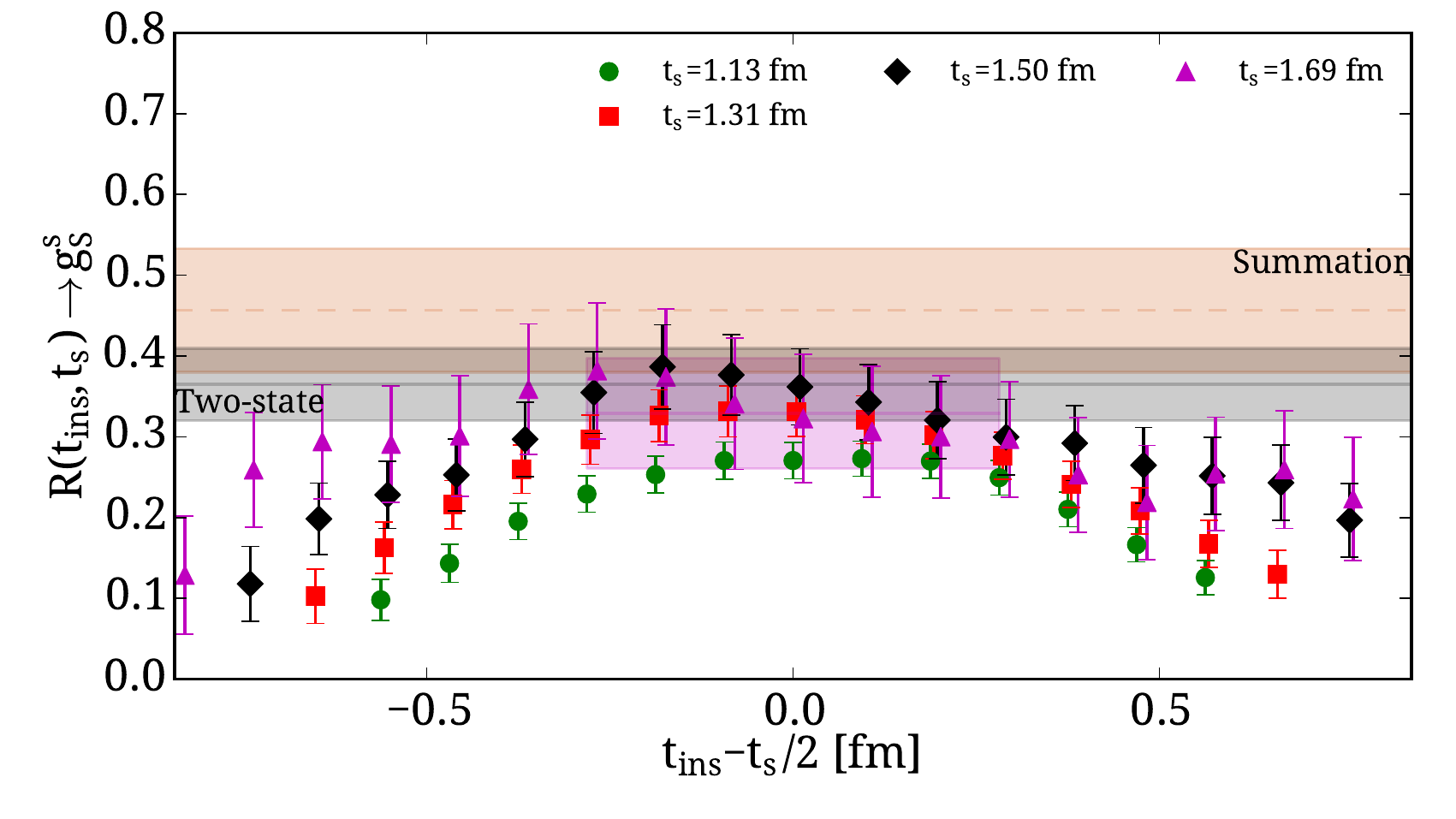}
\end{minipage}\hfill
\begin{minipage}{8.5cm}
\center
\includegraphics[width=\textwidth]{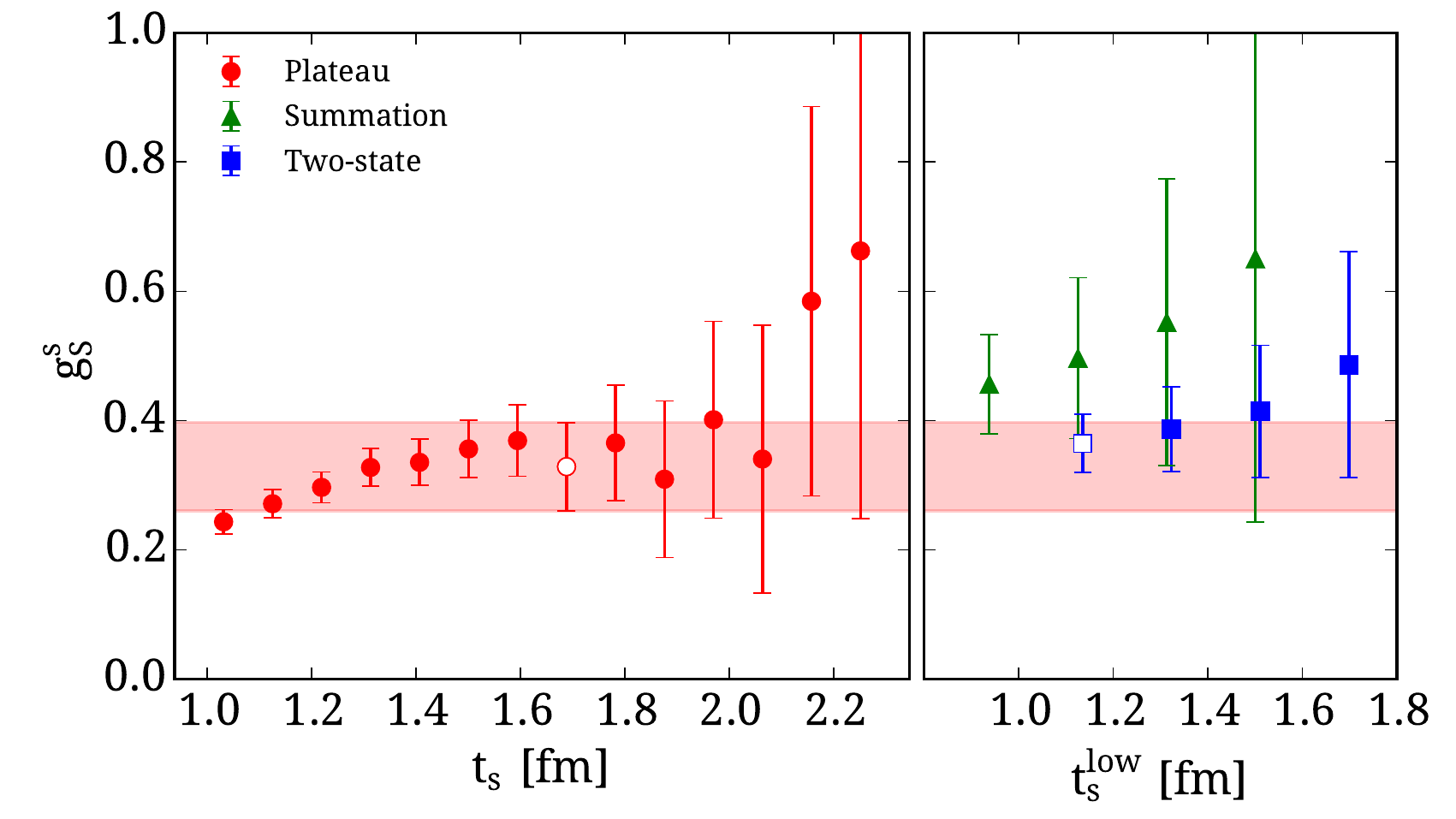}
\end{minipage}
\begin{minipage}{8.5cm}
\center
\includegraphics[width=\textwidth]{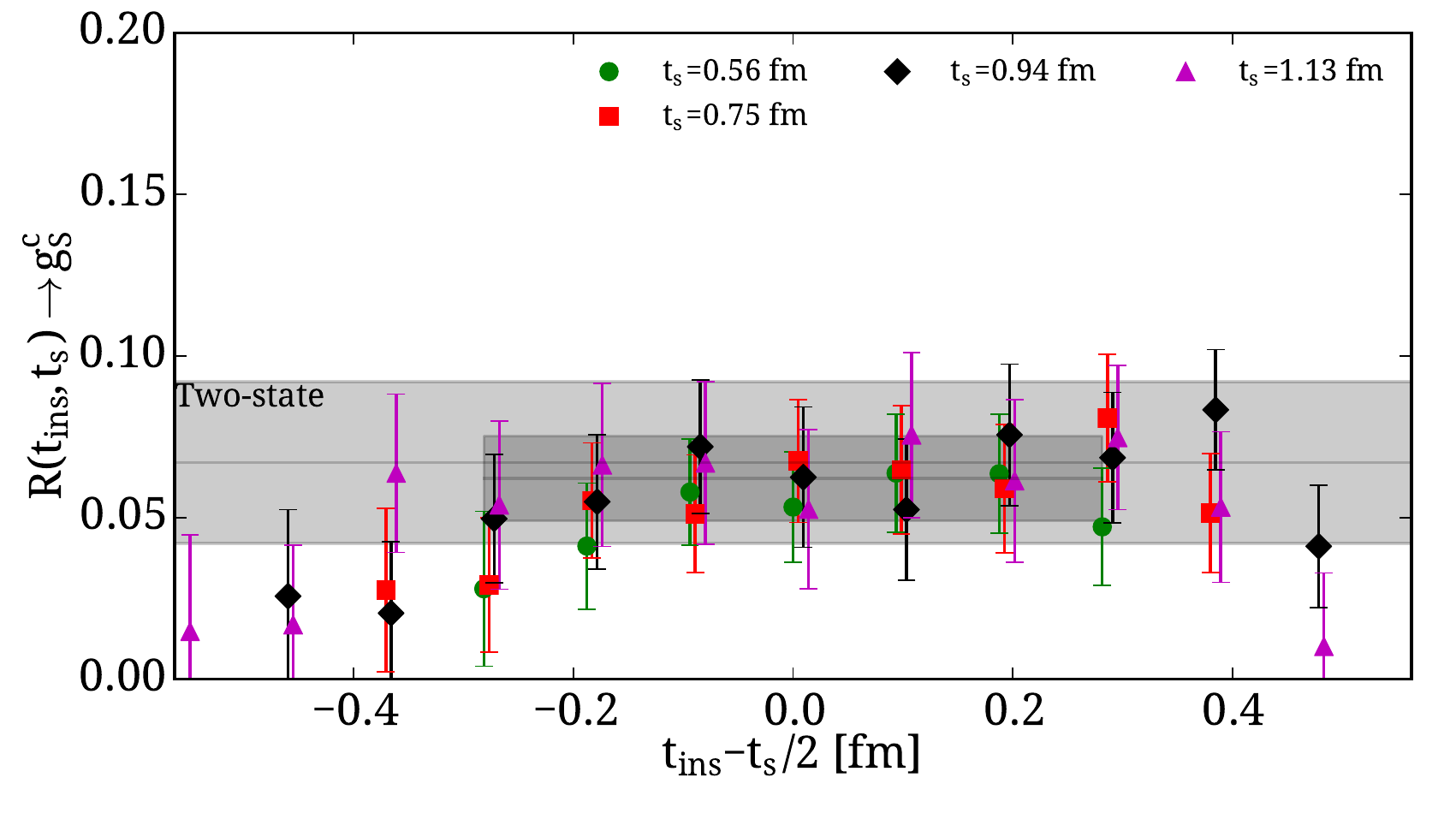}
\end{minipage}\hfill
\begin{minipage}{8.5cm}
\center
\includegraphics[width=\textwidth]{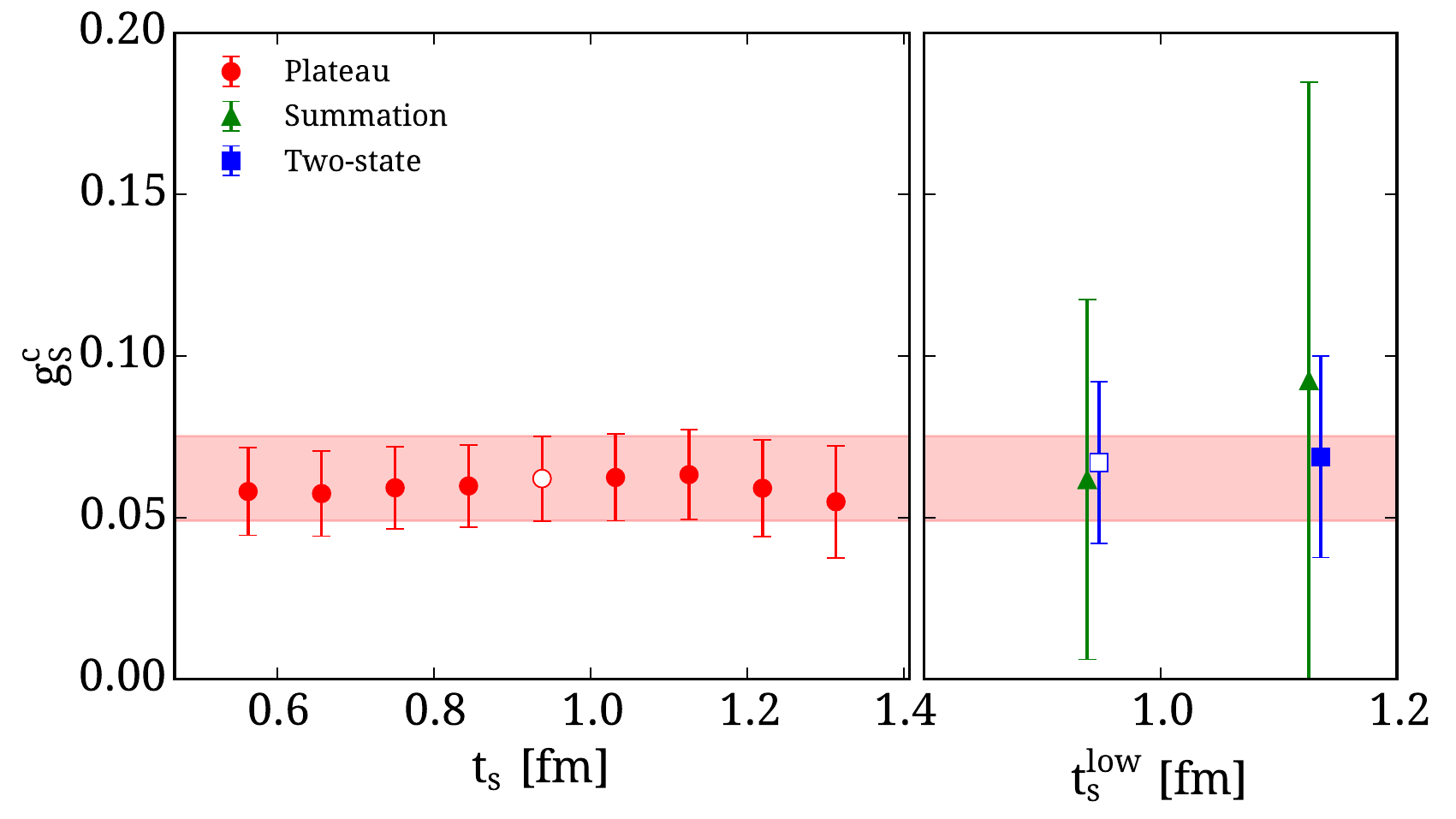}
\end{minipage}
\caption{The  strange (top) and charm (bottom) scalar charges.  The notation is as in~\fig{fig:gS_light_conn}. The various values of $t_s$ shown for the plateau method for each observable are listed in the corresponding legend of the plots.}
\label{fig:gS_sc}
\end{figure}

The results on the strange and charm scalar charges are purely disconnected and the ratios from which $g_S^s$ and $g_S^c$ are extracted are shown in~\fig{fig:gS_sc}. A first observation is that both $g_S^s$ and $g_S^c$ are non-zero and smaller than the disconnected contributions coming from the light quark loops, as expected. We find that the strange scalar charge is about $5\%$ of the isoscalar connected scalar charge, whereas the charm contribution is two orders of magnitude smaller. The source-sink separation $t_s=1.69$~fm at which the ratio for $g_S^s$ converges is comparable to the light charges. For $g_S^c$ the relative errors are larger and the plateau values overlap at smaller sink-source time separations. We choose the plateau value $t_s=0.94$~fm as our final estimate for $g_S^c$ with the systematic error due  to excited state effects as the difference between this and the value from the  two-state fit shown in~\fig{fig:gS_sc}.

In~\tbl{Table:gS_values} we collect the values for the light, strange and charm scalar charges from the plateau fits for all $t_s$. Also listed are the results from the summation method and the two-state fit that are shown in Figs.~\ref{fig:gS_light_conn},~\ref{fig:gS_light_disc} and~\ref{fig:gS_sc} with the brown and gray bands, respectively. The final values for the scalar charge that we select from the plateau fits are listed in~\tbl{Table:gS_gT_final_values}, where apart from the statistical error shown in the first parenthesis, we show in the second parenthesis the systematic error estimated from the error in the determination of the renormalization functions $Z_P^{\overline{\rm MS}}$, as well a systematic error due to excited state effects in the third parenthesis, taken as the difference in the mean values between the plateau and two-state fit methods.
%
% Strange and Charm scalar charge
\begin{table}
\begin{center}
\renewcommand{\arraystretch}{1.4}
\renewcommand{\tabcolsep}{4.5pt}
\begin{tabular}{c|ccccc|c|c}
\hline
\multirow{2}{*}{Observable} & \multicolumn{5}{c|}{Plateau at each $t_s$} & Two-state & Summation    \\
                            &  0.94  &  1.13  &  1.31  &  1.50  &  1.69   &    fit   &    method     \\
\hline\hline
$g_S^{u-d}$         & 0.732(81)  & 1.180(144) & 1.656(271) & 0.930(252) & 0.927(380) & 1.134(217) & 0.584(691)    \\      
$g_S^{u+d}$ (conn.) & 6.129(118) & 7.158(215) & 7.495(416) & 8.014(349) & 8.221(520) & 8.353(404) & 8.937(1.045) \\
$g_S^{u+d}$ (disc.) & 0.033(58)  & 0.399(71)  & 0.697(103) & 1.013(159) & 1.249(257) & 1.303(245) & 2.009(468) \\
$g_S^s$             & 0.220(17)  & 0.271(22)  & 0.328(29)  &  0.356(44) & 0.329(68)  & 0.365(45)  & 0.456(77)  \\
\hline
\multirow{3}{*}{$g_S^c$} &  \multicolumn{5}{c|}{Plateau at each $t_s$} & Two-state & Summation  \\
                      & 0.56  &  0.75   &  0.94  &    1.13     &    & fit    &   method  \\\cline{2-8}
                     & 0.058(14)  & 0.059(13)  & 0.062(13)  &  0.063(14) &   & 0.067(25) &  0.062(56)  \\
%\hline\hline
%
%
%
%                         &  \multicolumn{5}{c|}{%
%							\begin{tabular}{cccc}
%							 0.56      &    0.75     &    0.94     &    1.13 \\\hline
%							 0.056(12)  & 0.059(12)  & 0.061(12)  &  0.064(13) \\
%							\end{tabular}} & 0.062(56)  & 0.0672(25) \\
%\hline\hline
%
% Christos: Different way of presenting $g_S^c$
%\hline\hline
%\multirow{3}{*}{$g_S^c$} & \multicolumn{5}{c|}{Plateau at each $t_s$} & Summation & two-state  \\
%                         & \multicolumn{5}{c|}{%
%							\begin{tabular}{cccc}
%							   0.56      &    0.75     &    0.94     &    1.13    \\ \hline
%							  0.056(12)  &  0.059(12)  &  0.061(12)  &  0.064(13) \\
%							\end{tabular}} & method & fit \\
%							
%							0.062(56)  & 0.0672(25) \\
\hline
\end{tabular}
\caption{Results for the nucleon scalar charges with their jackknife errors. In columns two to six we give the results using the plateau method  at $t_s = 0.94$, $1.13$, $1.31$, $1.50$ and $1.69$~fm for the light and strange charges, whereas for $g_S^c$ the separations $t_s = 0.56$, $0.75$, $0.94$ and $1.13$~fm are given. In the last two columns we list the results from the two-state fit and the summation method that are shown in Figs.~\ref{fig:gS_light_conn},~\ref{fig:gS_light_disc} and~\ref{fig:gS_sc} with the bands (brown and gray respectively). The final value we select for each observable is shown in~\tbl{Table:gS_gT_final_values}.}
\label{Table:gS_values}
\end{center}
\end{table}

\subsection{Nucleon tensor charge}

\begin{figure}[h]
\begin{minipage}{8.5cm}
\center
\includegraphics[width=\textwidth]{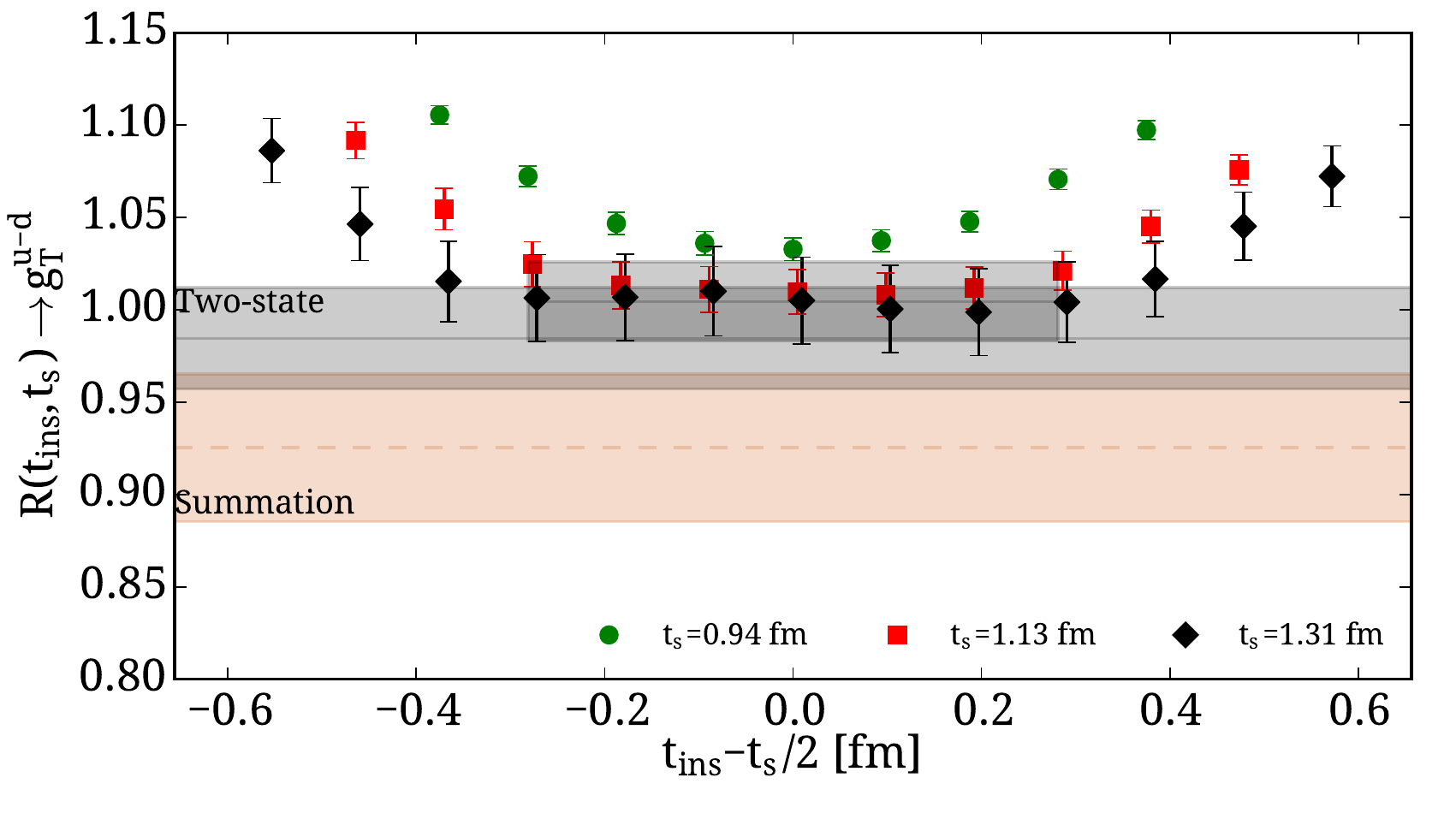}
\end{minipage}\hfill
\begin{minipage}{8.5cm}
\center
\includegraphics[width=\textwidth]{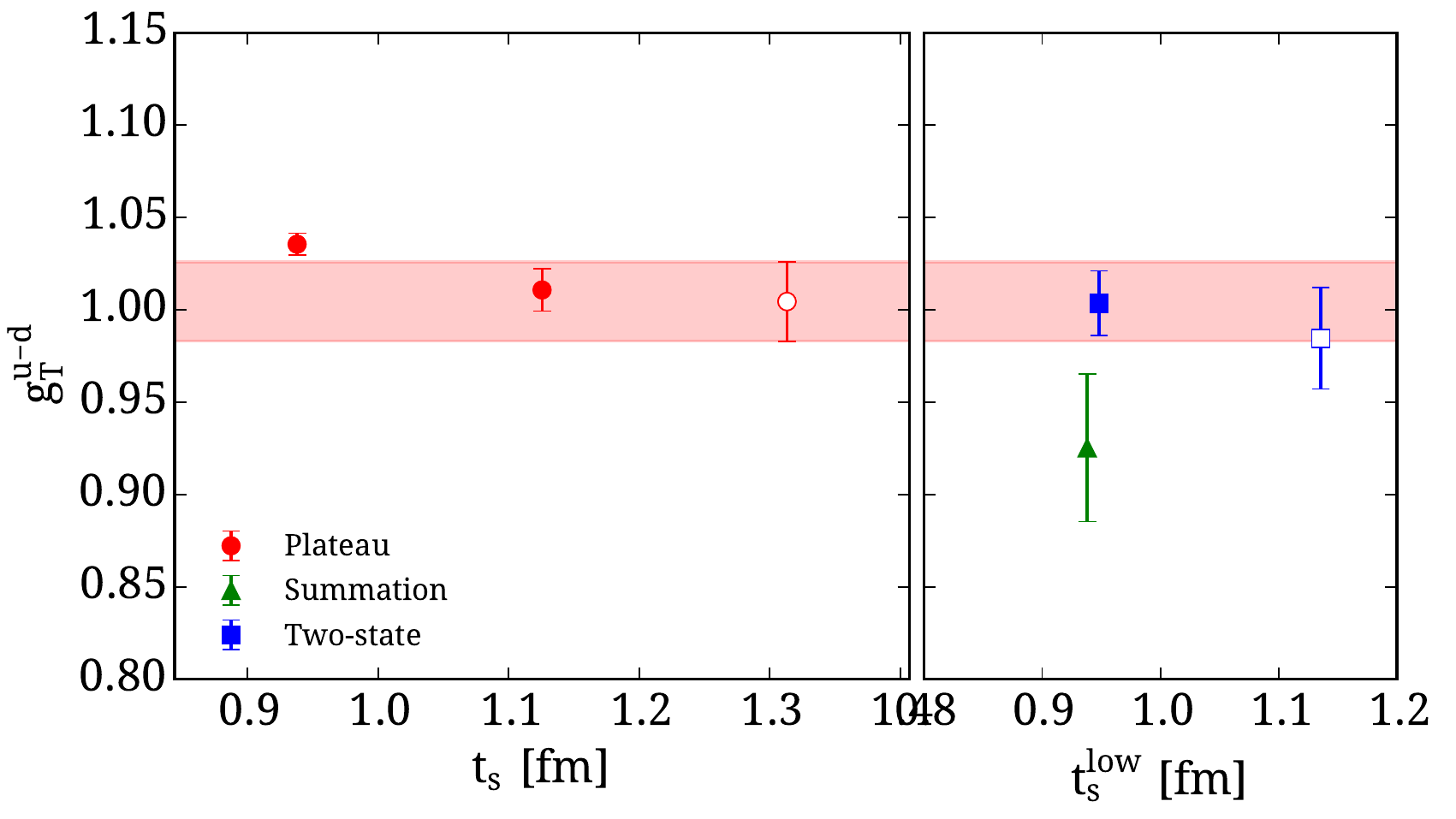}
\end{minipage}
\begin{minipage}{8.5cm}
\center
\includegraphics[width=\textwidth]{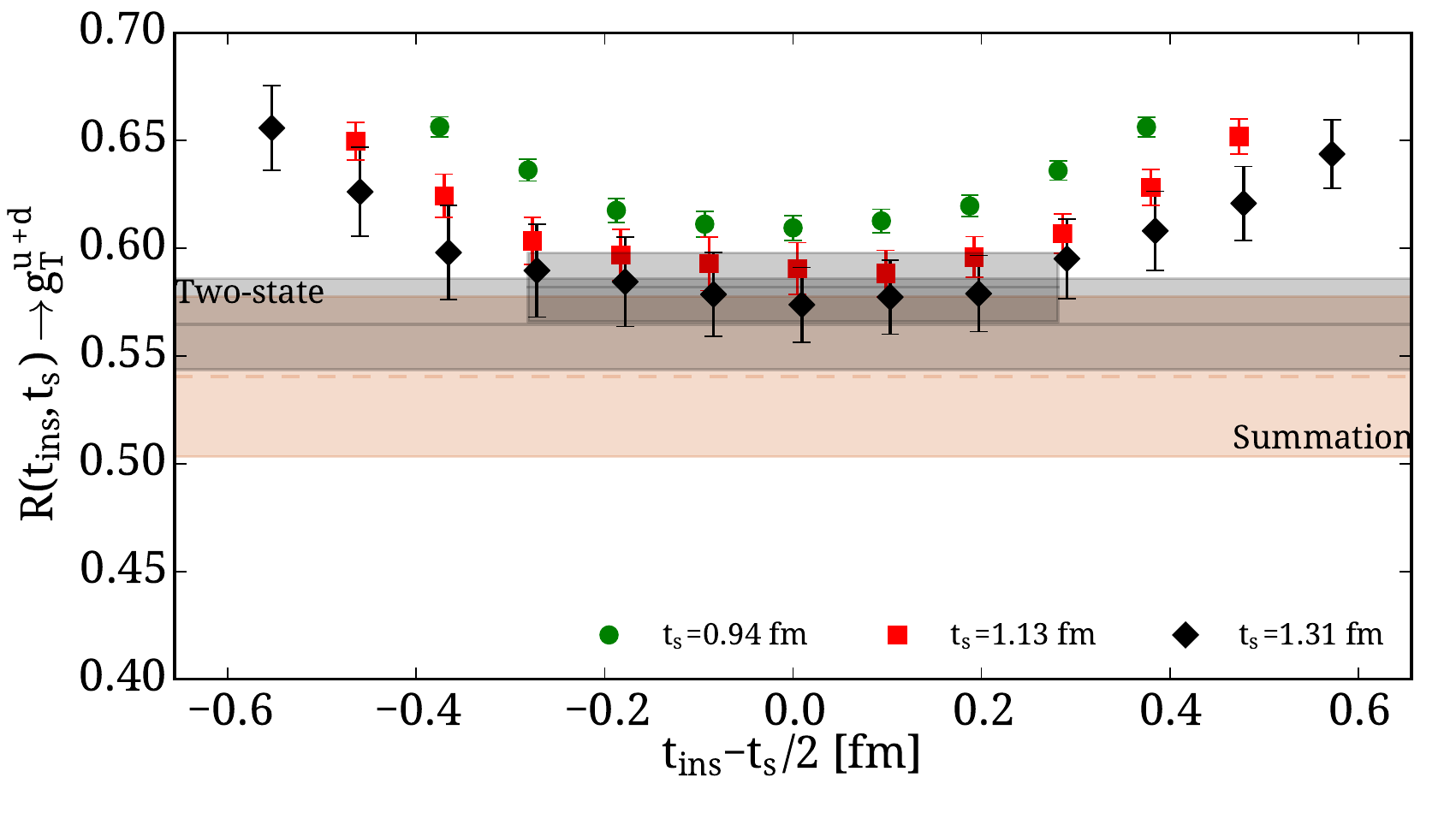}
\end{minipage}\hfill
\begin{minipage}{8.5cm}
\center
\includegraphics[width=\textwidth]{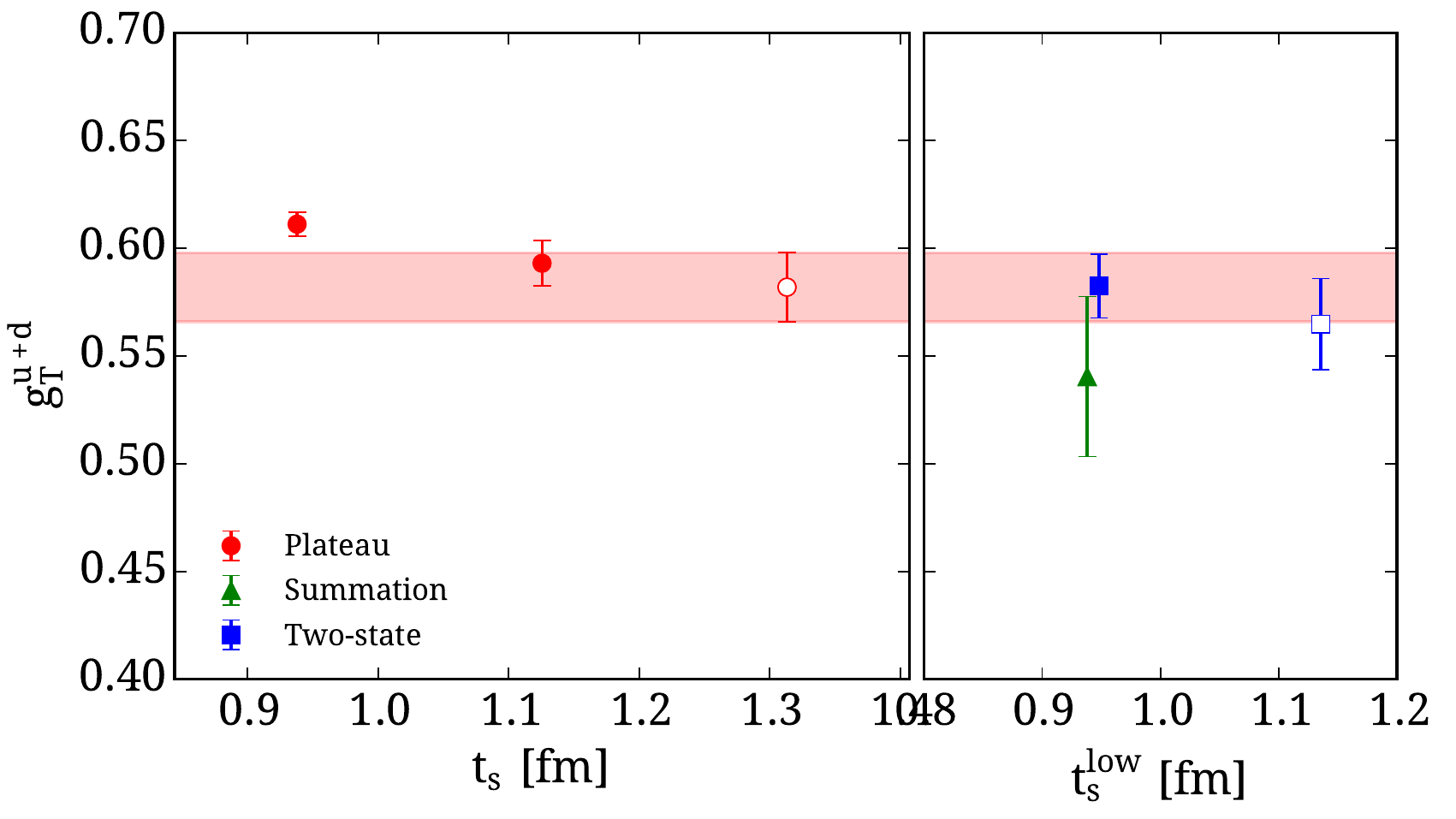}
\end{minipage}
\caption{The isovector (top) and connected isoscalar (bottom) nucleon tensor charge, following the notation of~\fig{fig:gS_light_conn}. The various values of $t_S$ shown for the plateau method are listed in the  legend of the plots.}
\label{fig:gT_light_conn}
\end{figure}

The results for the tensor charge are illustrated in this section in a similar manner to the discussion of the scalar charge . In~\fig{fig:gT_light_conn} we show the isovector and connected isoscalar tensor charges. From our previous study~\cite{Abdel-Rehim:2016won} we know that excited states effects are less severe for $g_T$. Indeed, a milder dependence on $t_s$ is confirmed also for this twisted mass ensemble  and the ratio appears to converge at $t_s=1.31$~fm for both the isovector and isoscalar quantities. This can be inferred also from the two-state fit and the summation method results where  consistent values are extracted  already for the lowest $t_s^{\rm low} = 0.94$~fm. Therefore, for the tensor charge the analysis is carried out only for three sink-source time separations.  

\begin{figure}[h]
\begin{minipage}{8.5cm}
\center
\includegraphics[width=\textwidth]{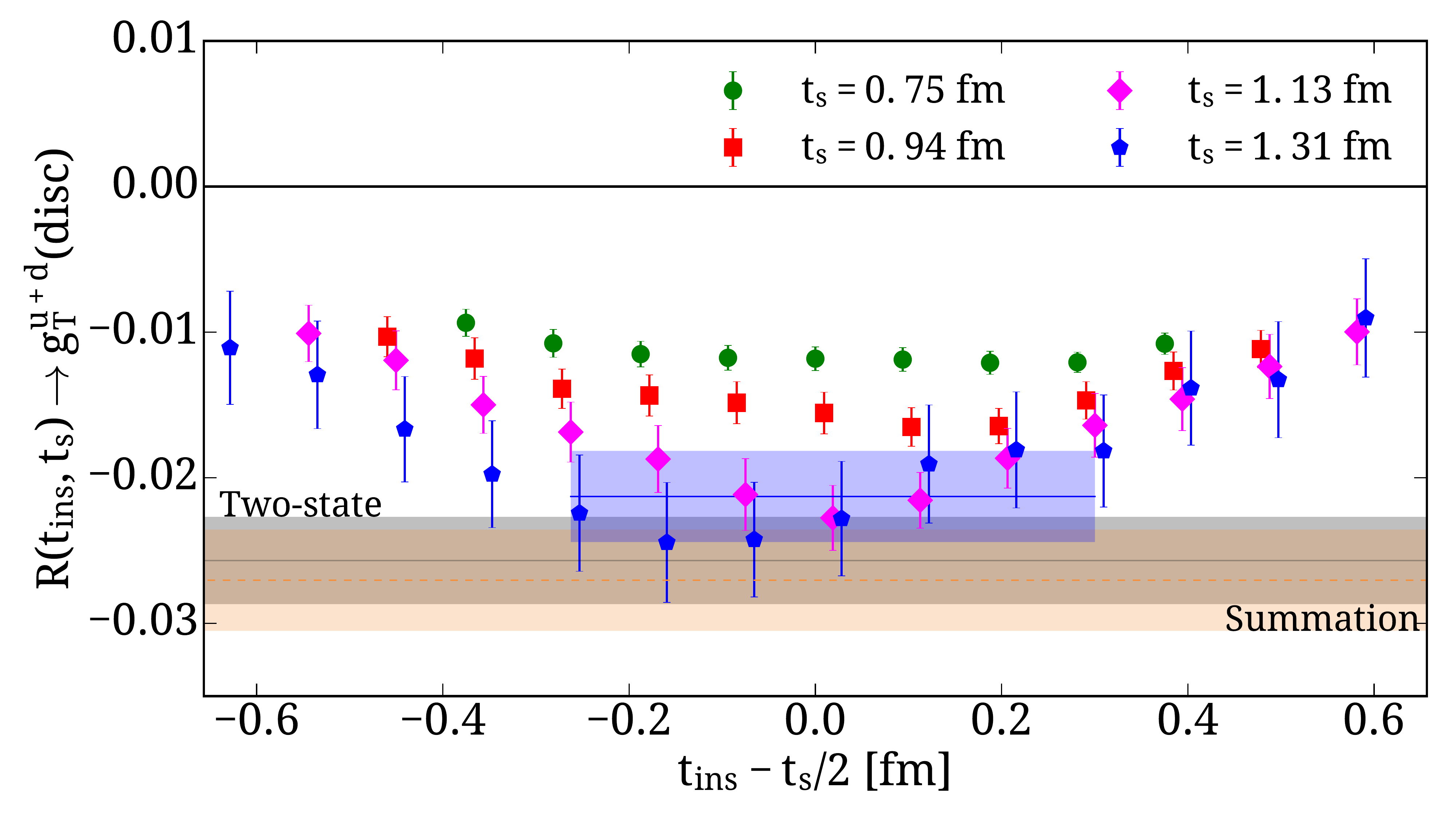}
\end{minipage}\hfill
\begin{minipage}{8.5cm}
\center
\includegraphics[width=\textwidth]{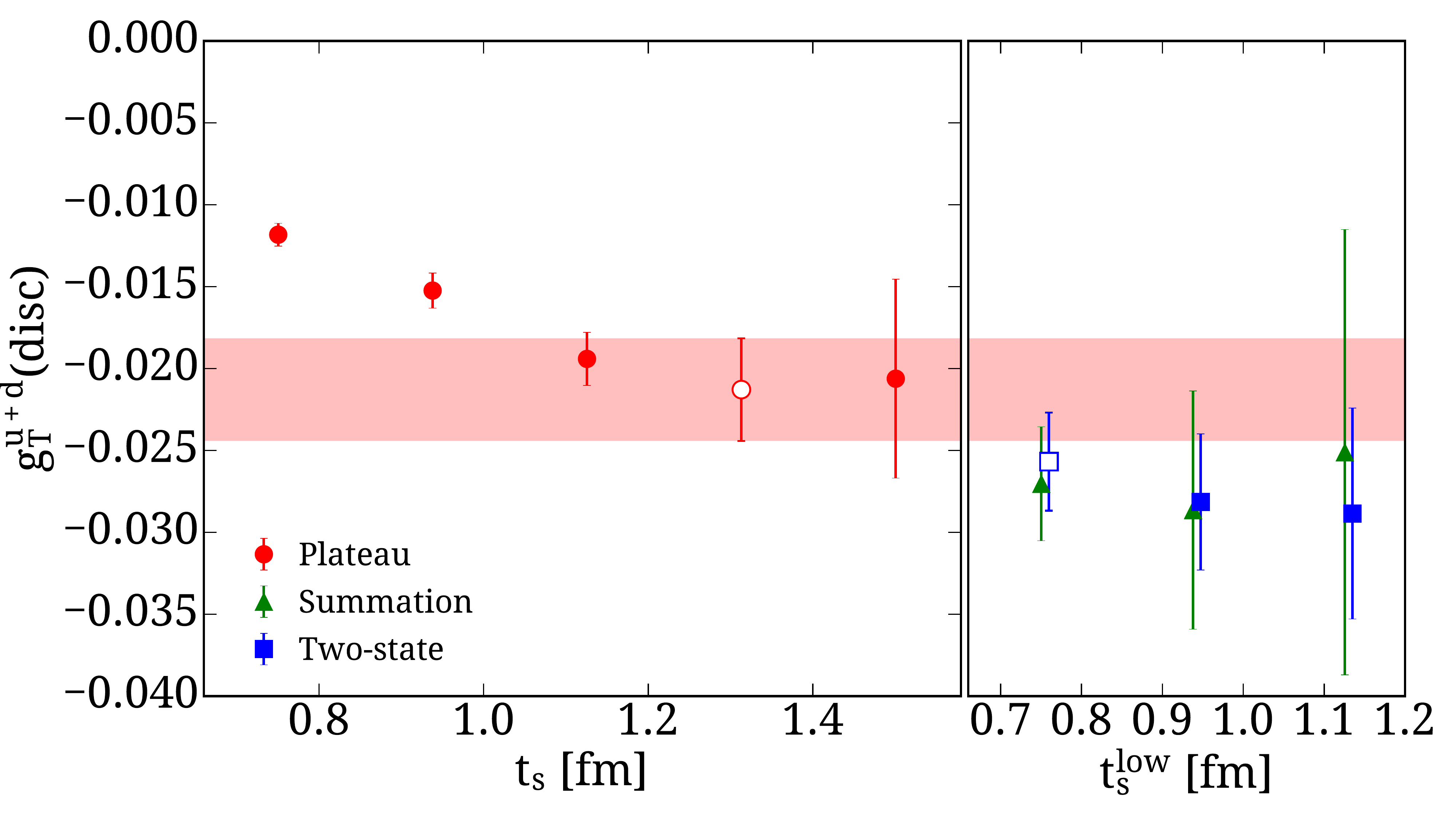}
\end{minipage}
\caption{Disconnected contributions to $g_T^{u+d}$. The notation is as in~\fig{fig:gS_light_conn}.  The various values of $t_s$ shown for the plateau method are listed in the legend of the plots.}
\label{fig:gT_light_disc}
\end{figure}

\begin{figure}[h]
\begin{minipage}{8.5cm}
\center
\includegraphics[width=\textwidth]{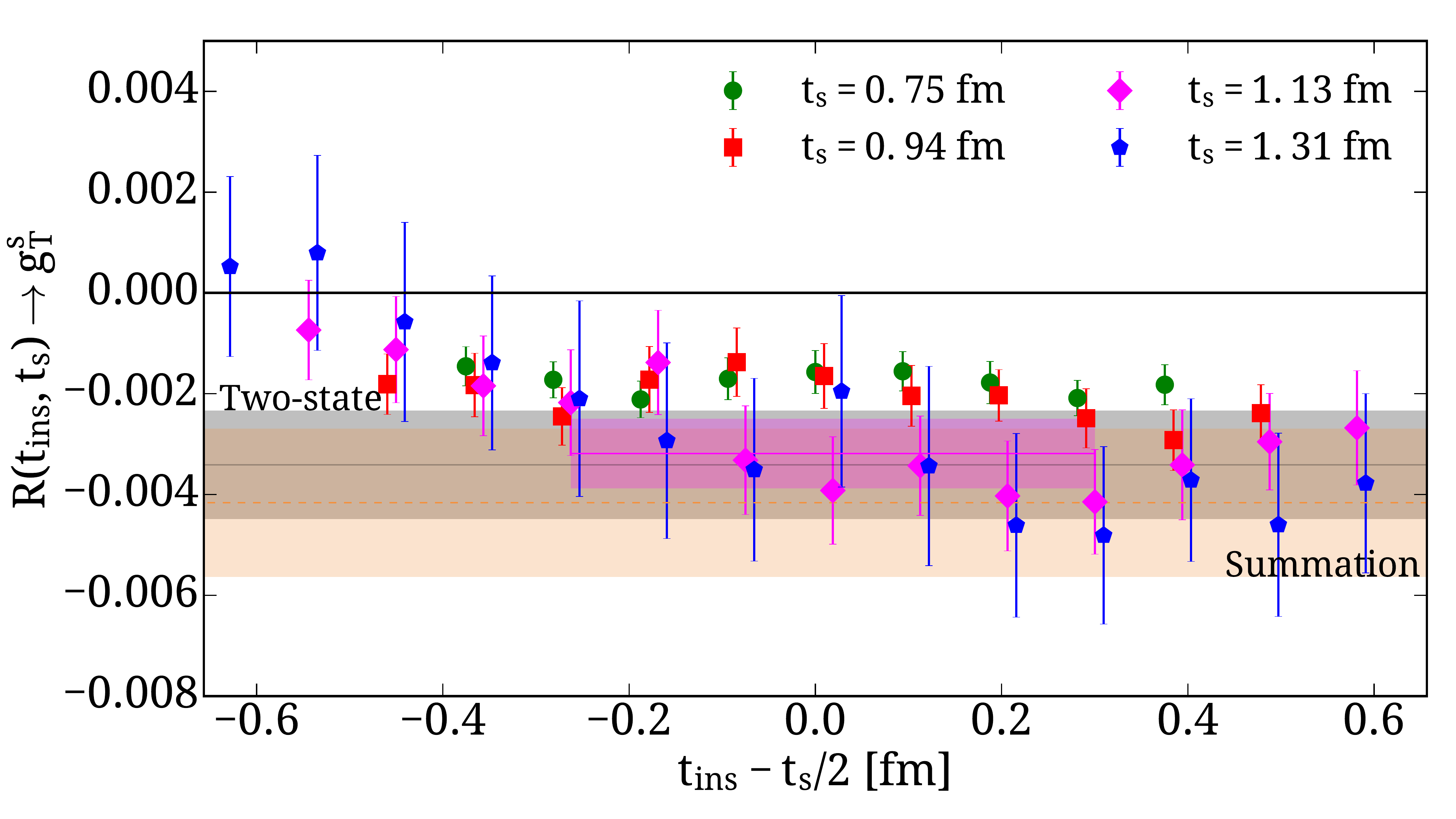}
\end{minipage}\hfill
\begin{minipage}{8.5cm}
\center
\includegraphics[width=\textwidth]{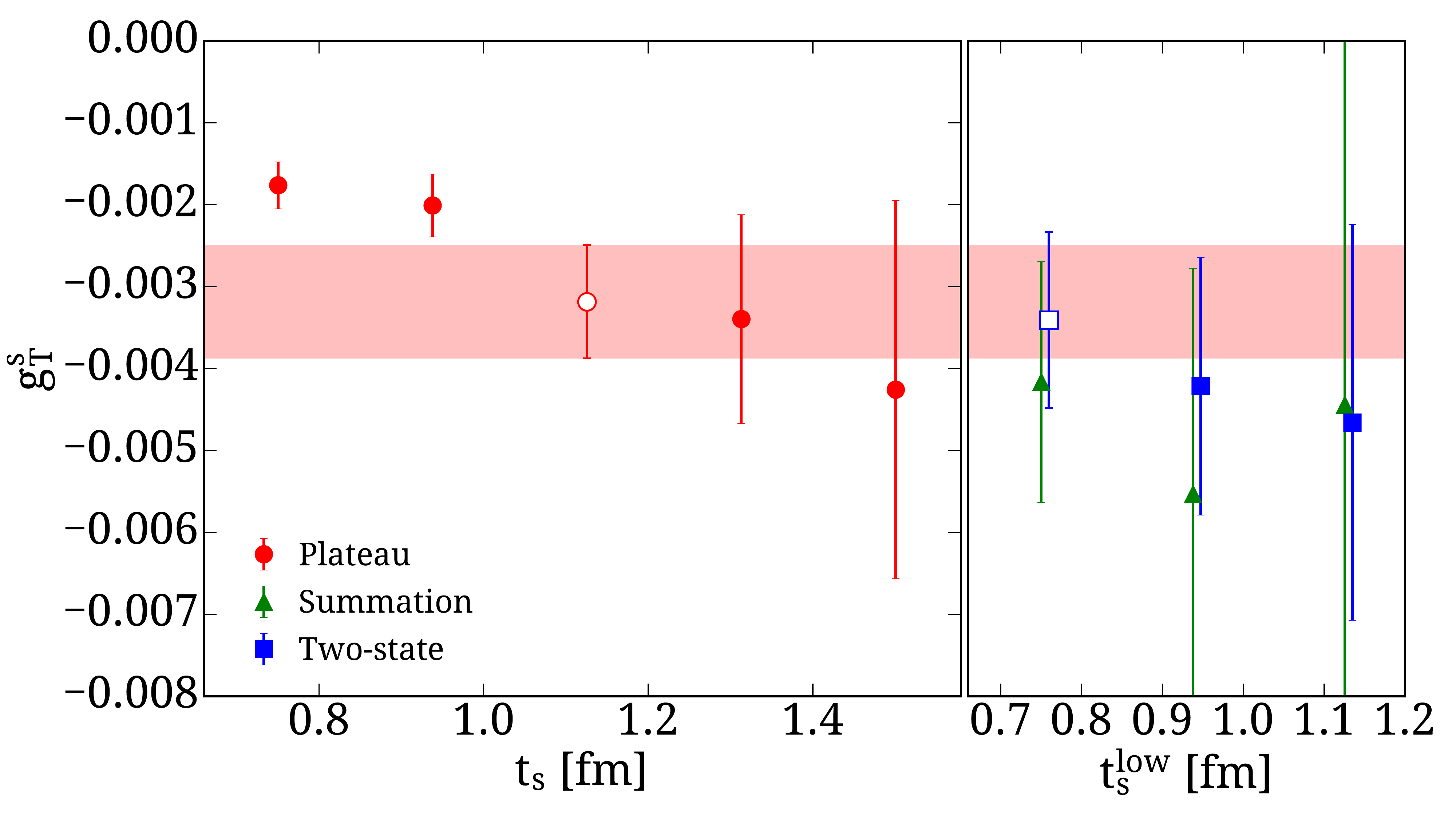}
\end{minipage}
\begin{minipage}{8.5cm}
\center
\includegraphics[width=\textwidth]{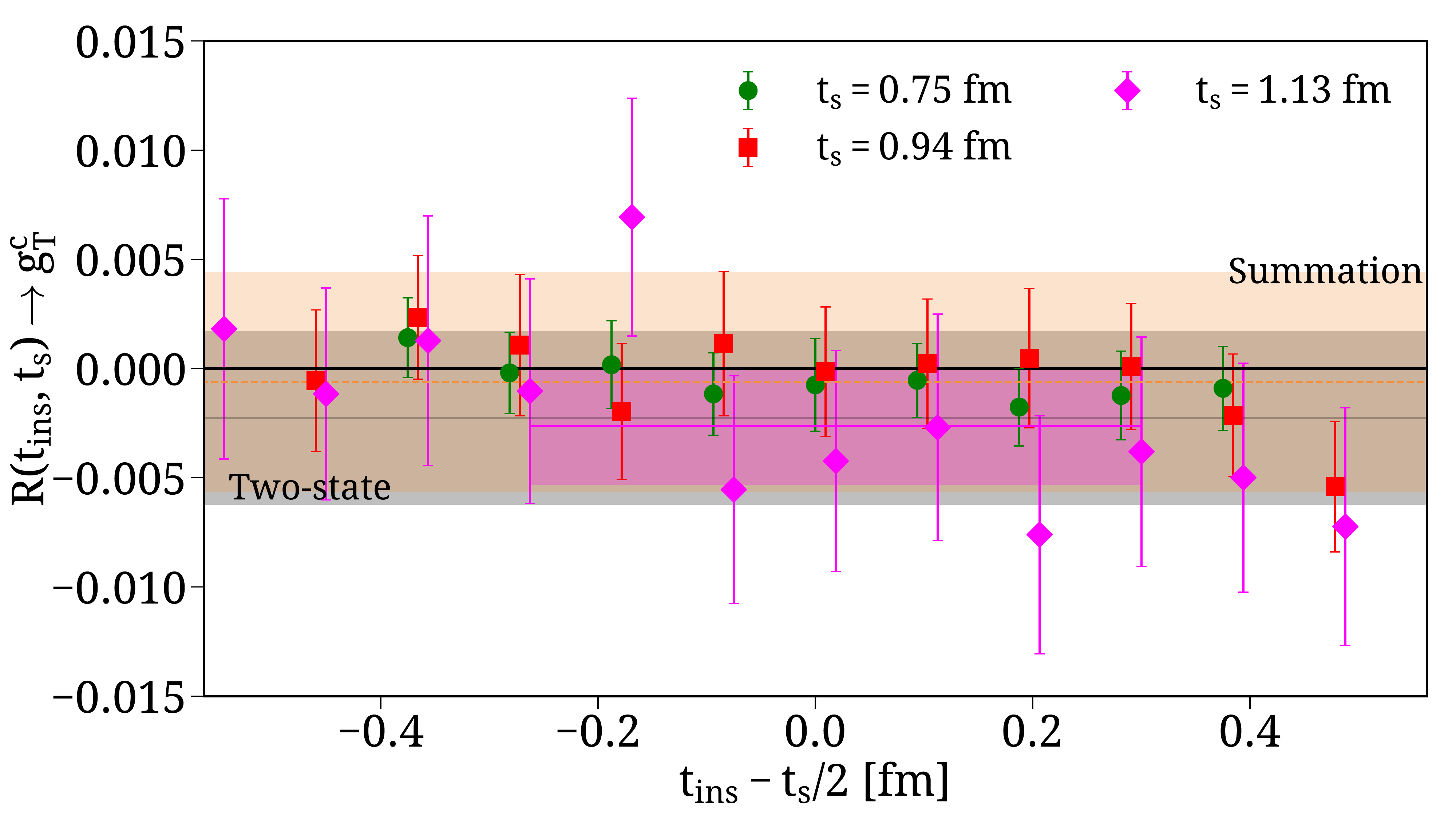}
\end{minipage}\hfill
\begin{minipage}{8.5cm}
\center
\includegraphics[width=\textwidth]{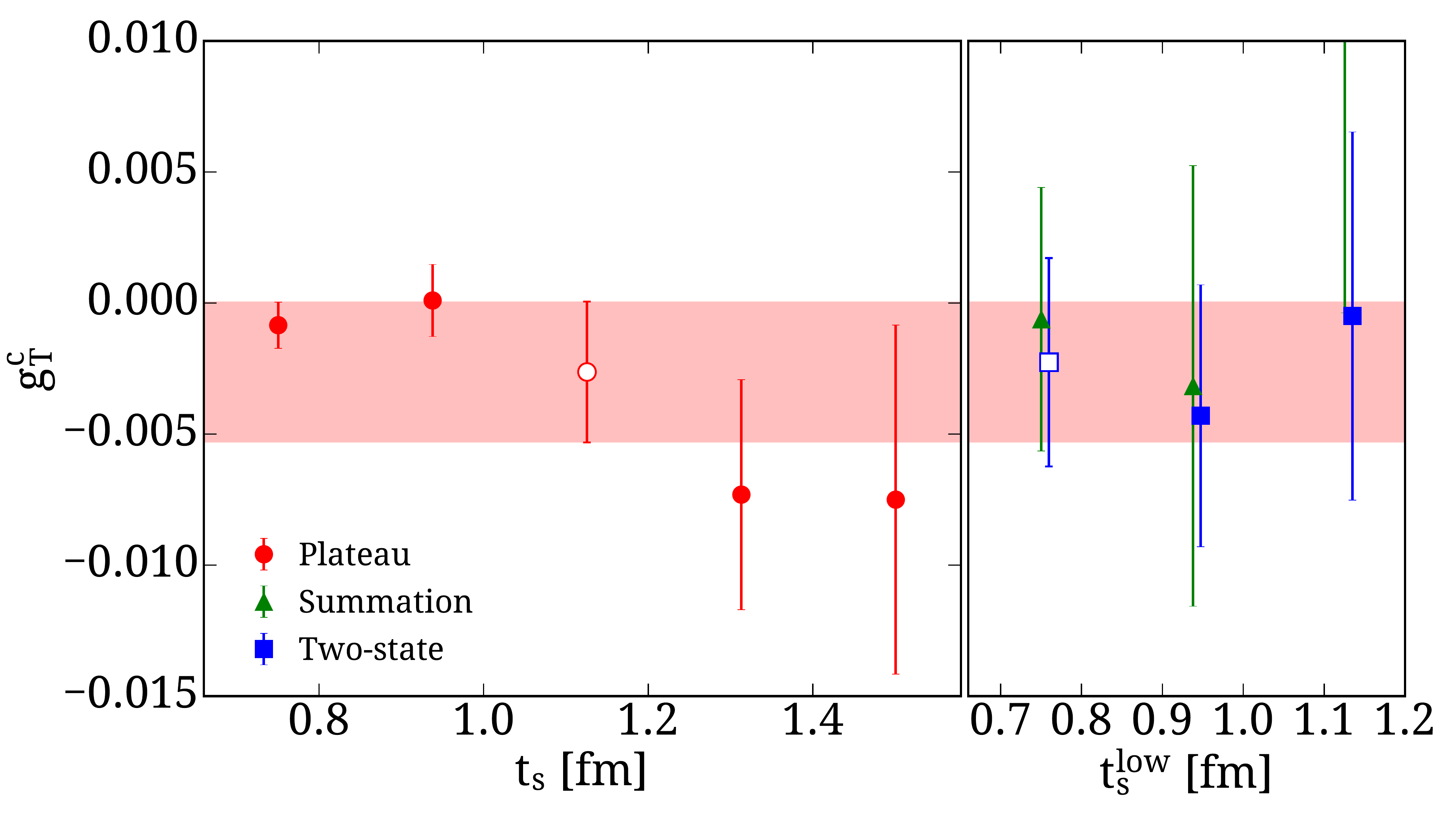}
\end{minipage}\caption{The purely disconnected strange (top) and charm (bottom) tensor charges. The notation is as in~\fig{fig:gS_light_conn}. The various sink times shown for the plateau method for each observable are listed in the corresponding legend of the plots.}
\label{fig:gT_sc}
\end{figure}

The disconnected contributions to $g_T^{u+d}$ are shown in~\fig{fig:gT_light_disc}. As can be seen, we obtain a clear non-zero negative value. The values extracted from fitting the plateau for different sink-source separation show convergence for $t_s=1.31$~fm as well as  agreement with the summation and two-state fit methods. We, thus, select the plateau value at this source-sink time separation as our final value for $g_T^{u+d}$. For the strange and charm tensor charges we obtain very small negative values, which are about 10\% of the disconnected light contributions, as demonstrated in~\fig{fig:gT_sc}. We select the plateau value at $t_s=1.13$~fm for both $g_T^s$ and $g_T^c$. Combining all disconnected contributions gives an upper bound of about 0.1\% when compared with the connected $g_T^{u+d}$.

%  The two-state fit yields very large
%  errors since the signal is too noisy to detect an excited state.
%  Given that the value at $t_s=0.75$~fm is consistent as we increase
%  further $t_s$ we take this as an upper bound for the small
%  disconnected contribution.  Any residual excited
%  state effects cannot be detected for these data and our conclusion
%  is that any disconnected contributions to the isoscalar tensor charge are less than 0.2\% of the connected. The same picture holds for the strange and charm tensor charges, as demonstrated in~\fig{fig:gT_sc}. These disconnected contributions have an upper bound of about 0.1\%.

% Connected IV and IS light tensor charge
%
%
% Disconnected IS (light) tensor charge
%
% Strange and Charm tensor charge

We tabulate the results of the connected tensor charge in~\tbl{Table:gT_values_conn} and of the disconnected contributions in~\tbl{Table:gT_values_disc}. 
%As can be seen, there are large uncertainties on the values from the summation method and the two-state fit for the disconnected not allowing a reliable determination of the systematic error, which we omit on these. In~\tbl{Table:gS_gT_final_values} we show the final values for the tensor charges selected from the plateau fits.
As with the scalar charges, the error in the first parenthesis is statistical, in the second parenthesis the systematic error due to the error in the determination of the renormalization functions $Z_T^{\overline{\rm MS}}$ and in the third parenthesis the systematic error due to excited state effects, taken as the difference in the mean values between the plateau and two-state fit methods.
%
% Table for light connected gT
\begin{table}
\begin{center}
\renewcommand{\arraystretch}{1.4}
\renewcommand{\tabcolsep}{4.5pt}
\begin{tabular}{c|ccc|c|c}
\hline
\multirow{2}{*}{Observable} & \multicolumn{3}{c|}{Plateau value } & Two-state & Summation  \\
                   $t_s$~(fm)         &    0.94    &    1.13     &    1.31         &  fit   &    method     \\
\hline\hline
$g_T^{u-d}$         & 1.036(6) & 1.011(11) & 1.004(21) & 0.985(27) & 0.925(40) \\      
$g_T^{u+d}$ (conn.) & 0.611(5) & 0.593(10) & 0.582(16) & 0.565(21) & 0.540(37) \\
\hline\hline
\end{tabular}
\caption{Results for the nucleon $g_T^{u-d}$ and connected $g_T^{u+d}$ with their jackknife errors. The results using the plateau method are shown at $t_s = 0.94$, $1.13$ and $1.31$~fm. In the last two columns we list the results from the summation method and the two-state fit that are shown in~\fig{fig:gT_light_conn} with the brown and gray bands respectively. The final value we select for each observable is shown in~\tbl{Table:gS_gT_final_values}.}
\label{Table:gT_values_conn}
\end{center}
\end{table}
%
% Table for disc. IS, strange and charm gT
\begin{table}
\begin{center}
\renewcommand{\arraystretch}{1.4}
\renewcommand{\tabcolsep}{4.5pt}
\begin{tabular}{c|ccccc|cc}
\hline
\multirow{2}{*}{Observable} & \multicolumn{5}{c|}{Plateau value}  & Two-state & Summation  \\
                          $t_s$~(fm)  & 0.75 & 0.84 & 0.94 & 1.13 & 1.31   &  fit   &    method    \
 \\
\hline\hline
$g_T^{u+d}$ (disc.) & -0.0118(7) & -0.0152(10) & -0.0194(16) & -0.0213(31) & -0.0206(61) & -0.0257(\
30) & -0.0270(35)  \\
$g_T^s$             & -0.00176(29) & -0.00200(38) & -0.00319(69) & -0.00340(127) & -0.00426(231)  &\
 -0.00341(108)  &  -0.00416(147)  \\
$g_T^c$             & -0.00085(88) & 0.00010(137) & -0.00263(269) & -0.00731(439) & -0.00750(666)  \
& -0.00226(398) & -0.00062(502)  \\
\hline
\end{tabular}
\caption{Comparison of results for the disconnected contributions to the nucleon
  tensor charge from the plateau method, the summation method and the
  two-state fit. The results  extracted using the plateau method are presented for source-sink time separations $t_s = 0.75$~fm to $t_s=1.31$~fm. The final value selected for each
  observable is shown in ~\tbl{Table:gS_gT_final_values}.}
\label{Table:gT_values_disc}
\end{center}
\end{table}

%
%
% Table with final values
\begin{table}
\begin{center}
\renewcommand{\arraystretch}{1.4}
\renewcommand{\tabcolsep}{4.5pt}
\begin{tabular}{ccccc}
\hline\hline
$g_S^{u-d}$ & $g_S^{u+d}$ (conn.) & $g_S^{u+d}$ (disc.) &  $g_S^s$  & $g_S^c$   \\
\hline
0.930(252)(48)(204)  &	8.221(520)(291)(132) &	1.249(257)(44)(54) & 0.329(68)(12)(36) & 0.062(13)(3)(5) \\
\hline\hline
$g_T^{u-d}$ & $g_T^{u+d}$ (conn.) & $g_T^{u+d}$ (disc.) & $g_T^s$     & $g_T^c$     \\
\hline
  1.004(21)(2)(0) &     0.582(16)(3)(1)  &       -0.0213(31)(1)(44)   & -0.00319(69)(2)(22) & -0.00263(269)(2)(37) \\
\hline
\end{tabular}
\caption{Final results of the nucleon's scalar and tensor charges, selected from the plateau fits. The error in the first parenthesis is the statistical, in the second parenthesis is a systematic error due to the error in the determination of the renormalization functions $Z_P^{\overline{\rm MS}}$ and $Z_T^{\overline{\rm MS}}$, whereas in the third parenthesis is the systematic error taking into account excited state contamination, and is taken as the difference in the mean values from the plateau and two-state fit methods.}
\label{Table:gS_gT_final_values}
\end{center}
\end{table}

%==========================================================================
%==========================================================================

\section{Comparison with other calculations}
\label{sec:comparisons}

In this section we compare the values of the nucleon scalar and tensor charges that we extract from the analysis presented here with a set of other recent lattice calculations as well as with results from phenomenology, when available.

As already mentioned, the scalar and tensor charges have received particular attention recently due to their implication in BSM physics. The isovector charges are the most studied in lattice QCD due to the absence of disconnected contributions. Whereas not all calculations are carried out using simulations directly at the physical pion mass, a number  of these results are extrapolated to the physical point, providing a comparison with our results that are computed directly at the physical point. The PNDME collaboration has recently presented results on the isovector scalar and tensor charges from nine $N_f=2+1+1$ ensembles using the highly-improved staggered quarks (HISQ) action produced by the MILC collaboration, at three values of the lattice spacing and a pion mass range of about $m_\pi = 140-315$~MeV~\cite{Bhattacharya:2016zcn}. Chiral extrapolations were performed and systematic uncertainties were studied. We also compare with the RQCD collaboration that obtained results from clover-improved fermions on eleven $N_f=2$ ensembles at three lattice spacings and several volumes, with a lowest pion mass of  $m_\pi=150$~MeV~\cite{Bali:2014nma}. The LHPC~\cite{Green:2012ej} has analyzed a number of $N_f=2+1$ ensembles of clover-improved fermions produced by the BMW collaboration, domain-wall-fermions (DWF) by the RBC and UKQCD collaborations, as well as a mixed action scheme which uses DWF valence quarks  on  Asqtad staggered sea quarks generated by the MILC collaboration  spanning a pion mass in the range of $m_\pi=149-356$~MeV. Both RQCD and LHPC performed chiral extrapolations and examined lattice artifacts.

\begin{figure}[h]
\begin{minipage}{8.5cm}
\center
\includegraphics[width=0.8\textwidth]{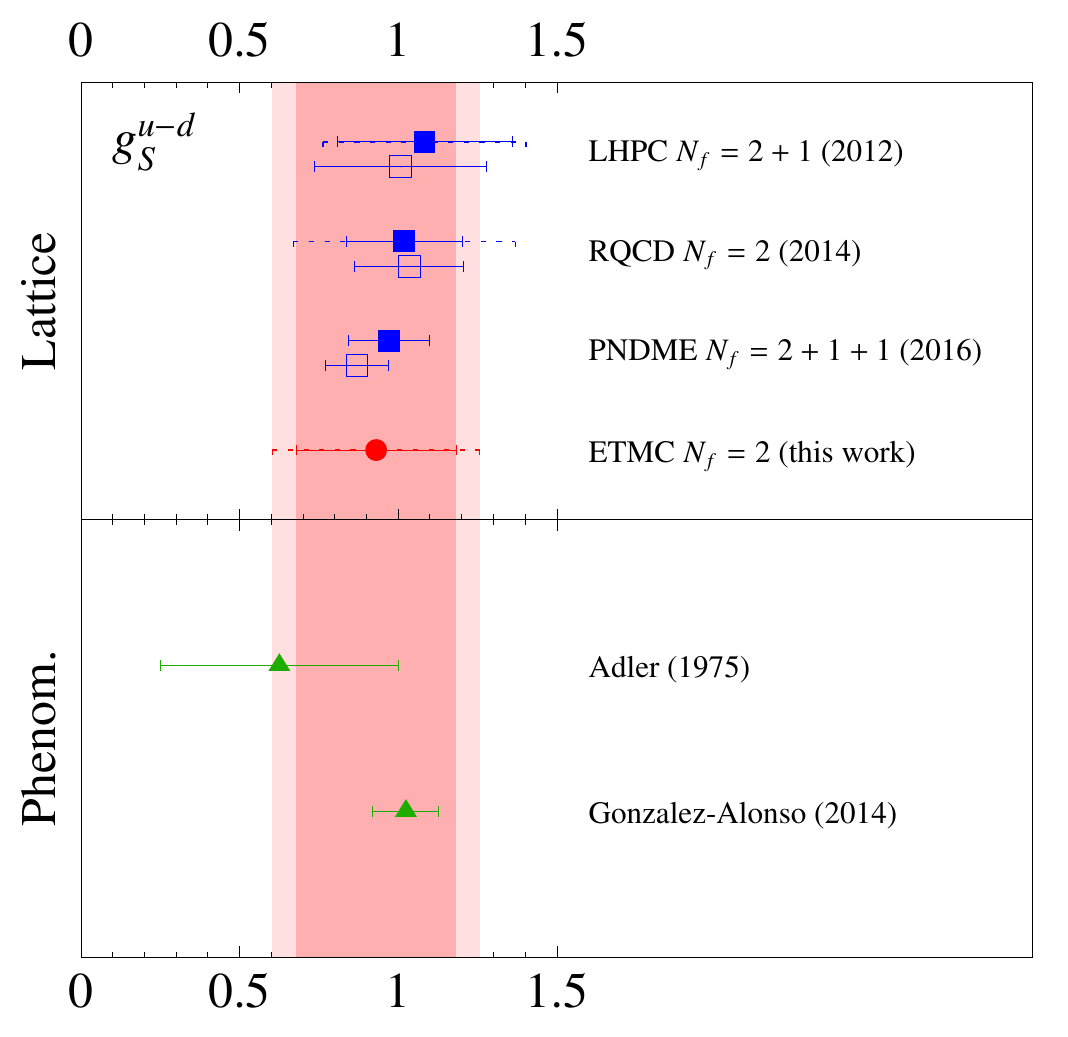}
\end{minipage}\hfill
\begin{minipage}{8.5cm}
\center
\includegraphics[width=0.8\textwidth]{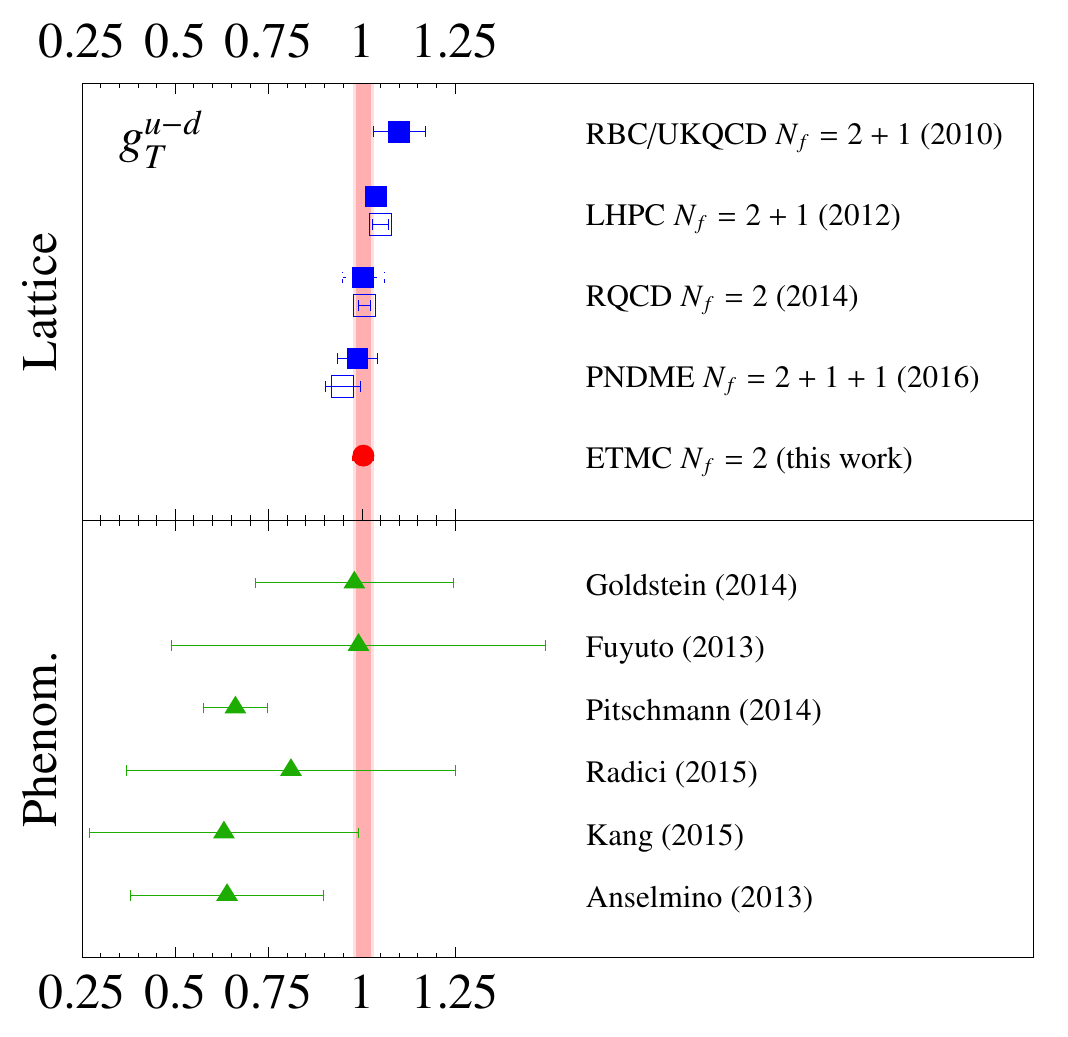}
\end{minipage}
\caption{Comparison of our results  (red circles) for $g_S^{u-d}$ (left) and $g_T^{u-d}$ (right) with a number of other recent lattice QCD results (blue squares) and with phenomenology (green triangles). With filled squares we denote extrapolated values at the physical pion mass, whereas with the open squares we show the lattice results from the various collaborations at their lowest pion mass, for the cases that $m_\pi^{\rm low}\le 150$~MeV. The solid error bars denote statistical errors whereas the dashed error bars show the statistical and systematic uncertainties added in quadrature.  The red vertical band showing our value and its errors is to help guide the eye.}
\label{fig:gS_gT_IV_world}
\end{figure}

In ~\fig{fig:gS_gT_IV_world} we compare our results on the isovector scalar charge with the values extrapolated at the physical point from the aforementioned collaborations, as well as with their value at their smallest pion mass if this is at or below $150$~MeV. In the same plot two results from phenomenological analyses are included. One is obtained by employing a quark model with spherically symmetric quark wave functions~\cite{Adler:1975he} to obtain an estimate for $g_S^{u-d}$. The second used a conserved vector current (CVC) relation $g_S/g_V = \delta M_N^{\rm QCD}/\delta m_q^{\rm QCD}$, where $\delta M_N^{\rm QCD}$ and $\delta m_q^{\rm QCD}$ are the mass differences of the proton and neutron and the up- and down-quarks, respectively, in pure QCD~\cite{Gonzalez-Alonso:2013ura,Abdel-Rehim:2016won}. As can be seen, there is a very good agreement among all lattice calculations as well as with the phenomenology results.

In~\fig{fig:gS_gT_IV_world} we also compare the value we extract for the isovector tensor charge  with the lattice calculations from PNDME, RQCD and LHPC, and additionally from the RBC/UKQCD collaboration using $N_f=2+1$ domain-wall-fermions at a pion mass range of $m_\pi=330-670$~MeV~\cite{Aoki:2010xg}. We furthermore include a number of phenomenology results from Refs.~\cite{Kang:2015msa,Goldstein:2014aja,Pitschmann:2014jxa,Anselmino:2013vqa,Radici:2015mwa,Fuyuto:2013gla}. The lattice QCD results are very accurate and show an excellent agreement among them. Their errors are noticeably smaller as compared to the phenomenological results, illustrating the important input that lattice QCD is currently providing.
%{\bf Christos: Removed this paragraph: Also quite interesting is the very accurate result $g_T^{u-d} = 0.61_{-0.017}^{+0.017}$ that JLab published recently~\cite{Ye:2016prn}, obtained by simulating pseudodata from the Solenoidal Large Intensity Device (SoLID) experiment and studying their impact on the determination of the tensor charge. This result, while featuring comparable errors to the existing lattice QCD calculations, its value is about $40\%$ smaller and agrees with most phenomenology results. {\bf Christos: can we give a comment here? E.g. errors are still too large, need to better understand systematics, work on this is ongoing...} }
%
% Isovector charges, world

\begin{figure}[h]
\begin{minipage}{8.5cm}
\center
\includegraphics[width=\textwidth]{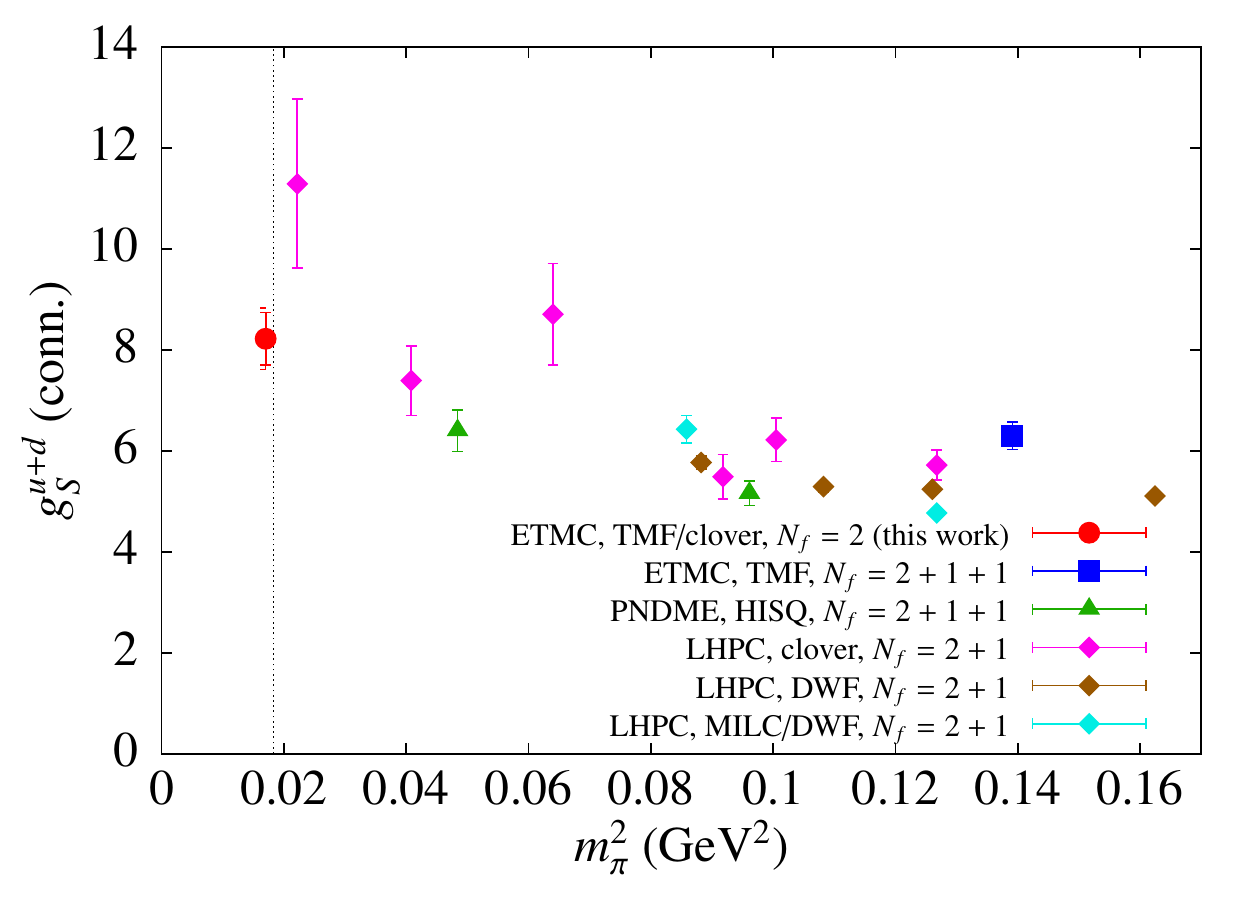}
\end{minipage}\hfill
\begin{minipage}{8.5cm}
\center
\includegraphics[width=\textwidth]{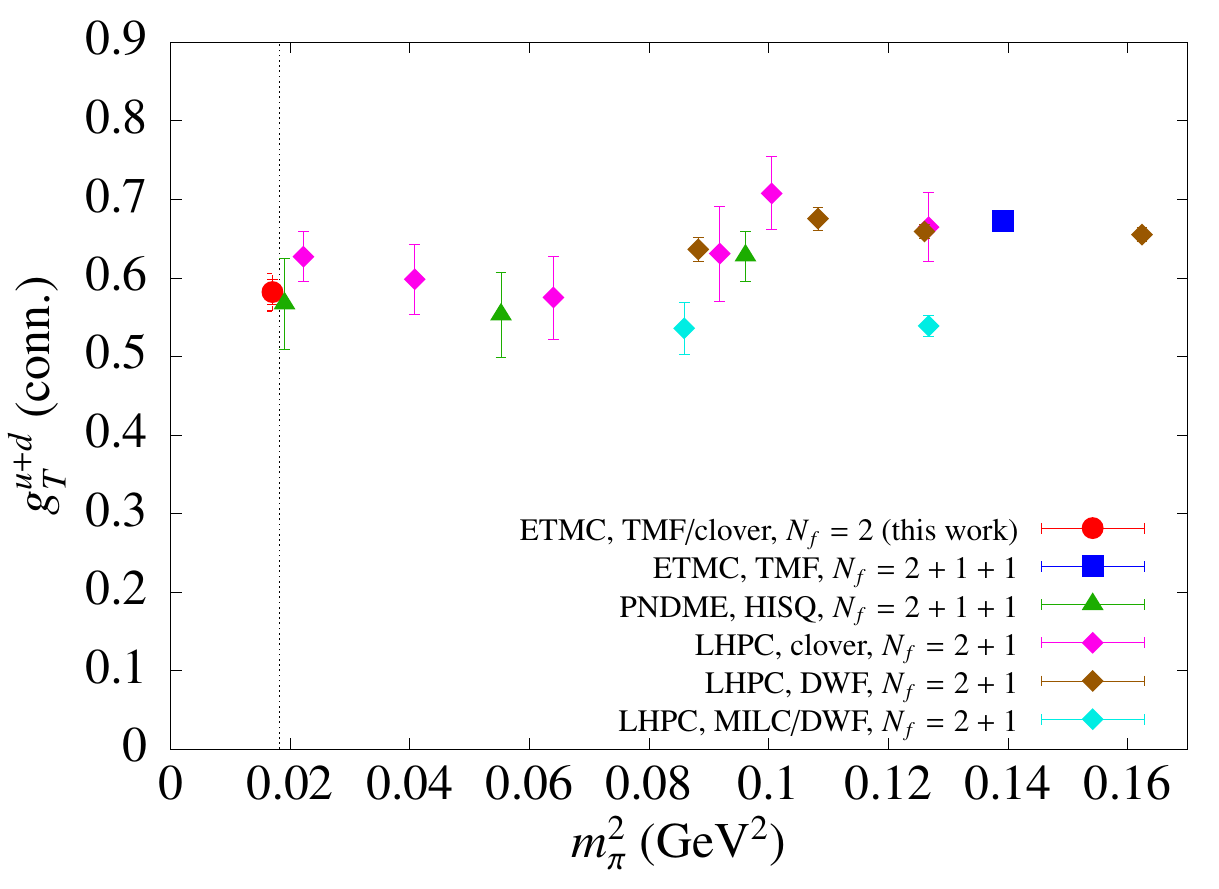}
\end{minipage}
\caption{Comparison of our results using the physical ensemble (red circles) for the connected $g_S^{u+d}$ (left) and $g_T^{u+d}$ (right) with lattice results from the ETMC using TMF fermions on $N_f=2+1+1$ gauge configurations~\cite{Abdel-Rehim:2013wlz} (blue square), the PNDME using $N_f=2+1+1$ staggered fermions (green triangles) from Ref.~\cite{Bhattacharya:2013ehc} for $g_S^{u+d}$ and Ref.~\cite{Bhattacharya:2016zcn} for $g_T^{u+d}$ and from the LHPC~\cite{Green:2012ej}, using $N_f=2+1$ clover fermions, domain-wall fermions and a mixed action approach (magenta, brown and light blue diamonds, respectively). The solid error bars in our results denote statistical errors whereas the dashed error bars show the statistical and systematic uncertainties added in quadrature.}
\label{fig:gS_gT_IS_conn_world}
\end{figure}

In~\fig{fig:gS_gT_IS_conn_world} we compare our results for the connected parts of the isoscalar scalar and tensor charges with selected results from the PNDME and the LHPC collaborations at various pion masses, using the lattice ensembles described previously, as well as with TMF from a gauge ensemble at $m_\pi=373$~MeV~\cite{Abdel-Rehim:2013wlz}. Regarding $g_S^{u+d}$, PNDME obtained results at two pion masses, $m_\pi=220$~MeV and $310$~MeV using clover valence quarks on  $N_f=2+1+1$ HISQ sea fermions~\cite{Bhattacharya:2013ehc} and performed a linear extrapolation to the physical point to obtain $g_S^{u+d} = 7.15(65)$, which  agrees with our value. The same group calculated $g_T^{u+d}$ on the nine gauge ensembles for which they obtained the isovector charges~\cite{Bhattacharya:2016zcn} and after performing a chiral extrapolation they obtain $g_T^{u+d}=0.598(33)$, which is in good agreement with the value extracted in this work. In general, there is agreement among lattice QCD for $g_S^{u+d}$ and $g_T^{u+d}$ over a range of pion masses. The tendency for lower values regarding $g_S^{u+d}$ and higher values regarding $g_T^{u+d}$ at heavier pion masses can be explained by the fact that older results have typically used smaller sink-source time separations. Since these quantities are affected by excited state contaminations that tend to decrease and increase their values, respectively, contributions from  excited states might explain the higher and lower values, respectively, obtained in more recent calculations.
%
% Connected Isoscalar charges, world

\begin{figure}[h]
\begin{minipage}{8.5cm}
\center
\includegraphics[width=\textwidth]{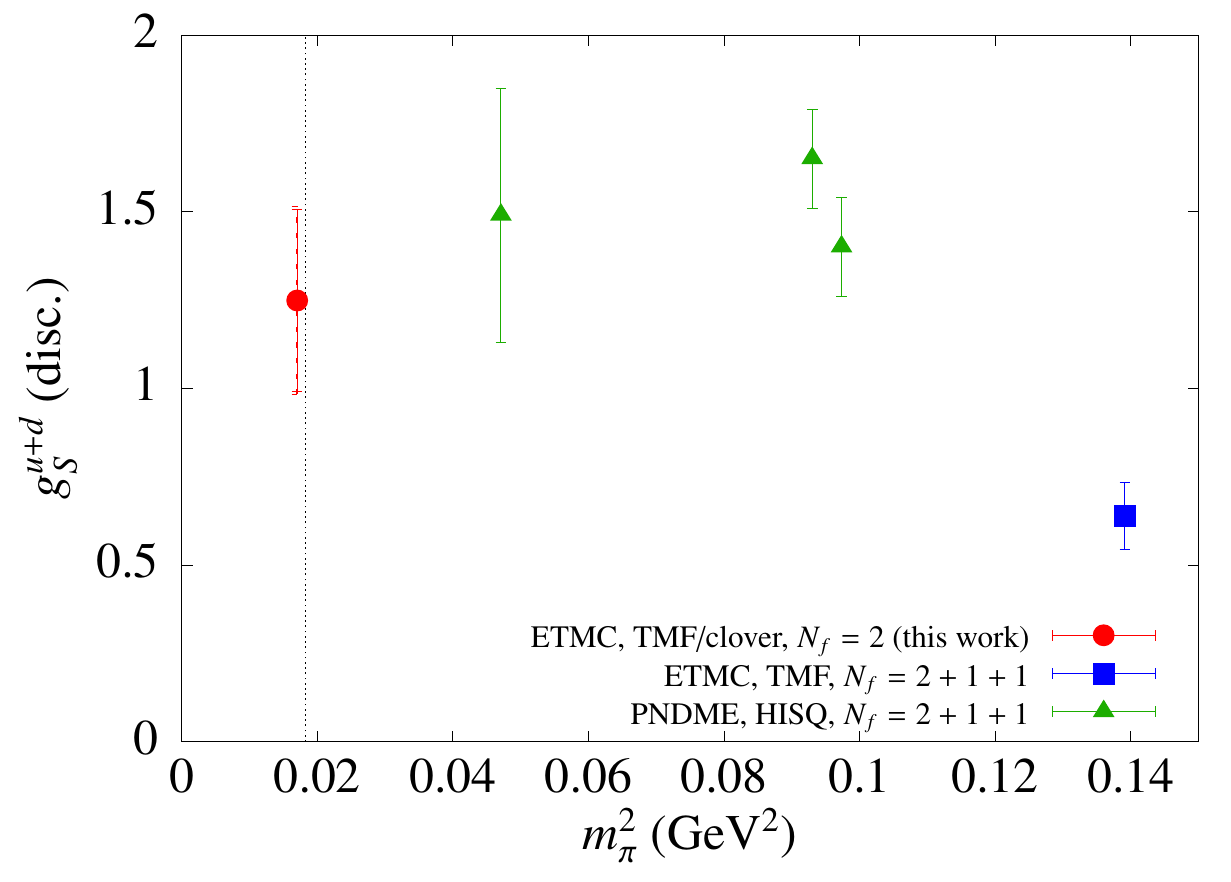}
\end{minipage}\hfill
\begin{minipage}{8.5cm}
\center
\includegraphics[width=\textwidth]{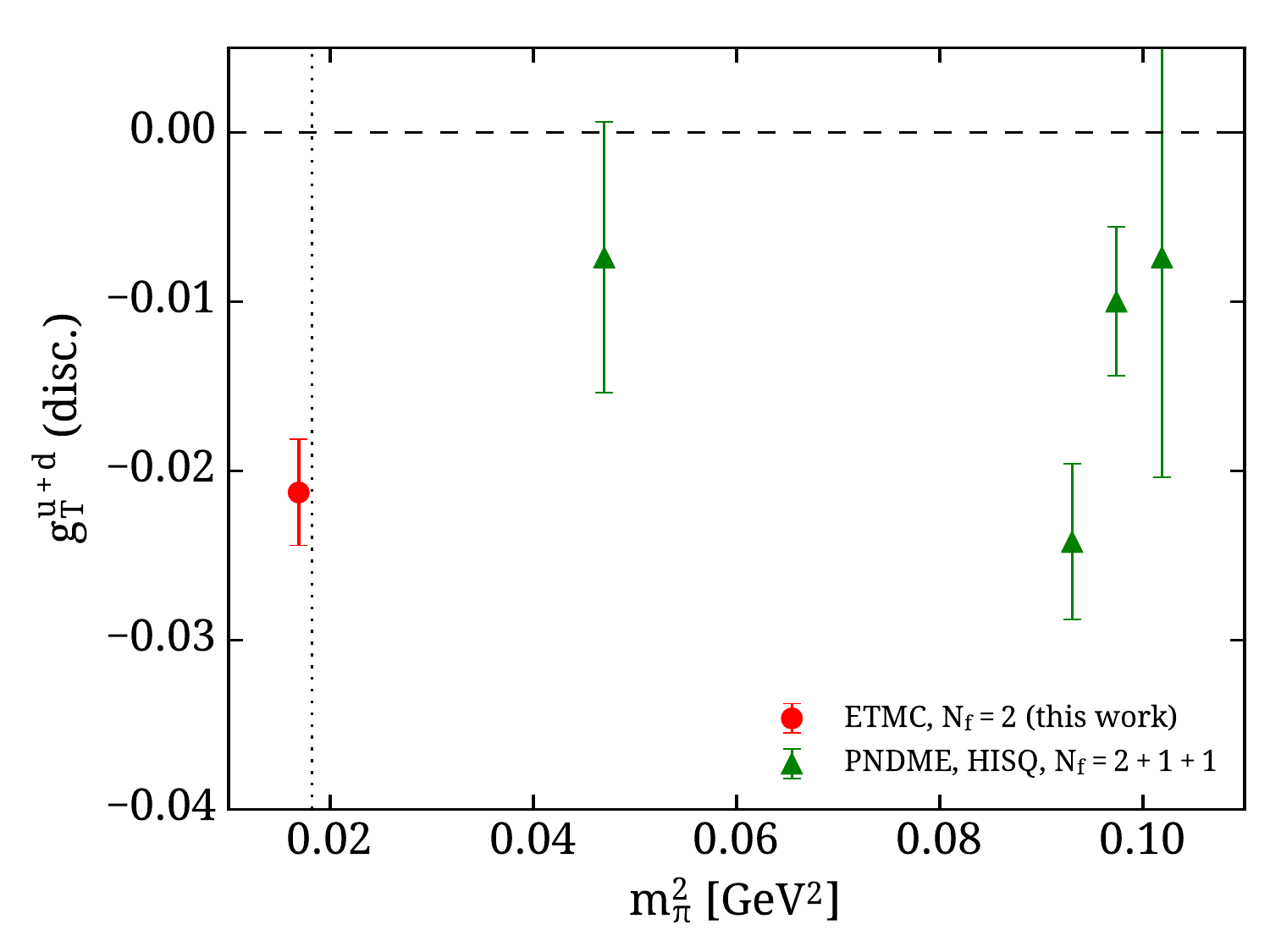}
\end{minipage}
\caption{Comparison of our results using the physical ensemble (red circles) for the disconnected contribution to $g_S^{u+d}$ (left) and $g_T^{u+d}$ (right) with lattice results from the ETMC using TMF fermions on $N_f=2+1+1$ gauge configurations~\cite{Abdel-Rehim:2013wlz} (blue square) and the PNDME collaboration, using clover valence fermions on a $N_f=2+1+1$ HISQ quark sea (green triangles) from Ref.~\cite{Bhattacharya:2015gma} for $g_S^{u+d}$ and Ref.~\cite{Bhattacharya:2015wna} for $g_T^{u+d}$. The solid error bars in our results denote statistical errors whereas the dashed error bars show the statistical and systematic uncertainties added in quadrature.}
\label{fig:gS_gT_IS_disc_world}
\end{figure}

\begin{figure}[h]
\begin{minipage}{8.5cm}
\center
\includegraphics[width=\textwidth]{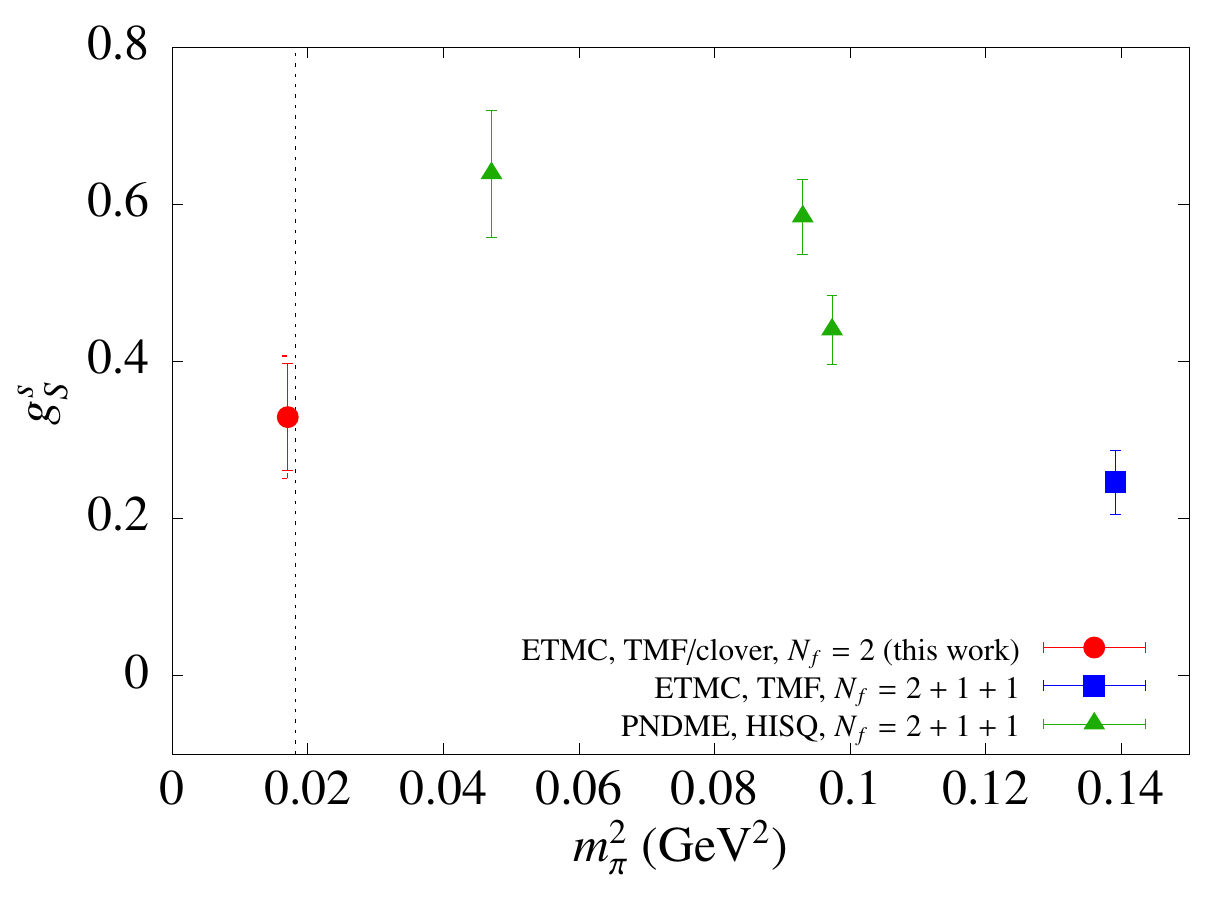}
\end{minipage}\hfill
\begin{minipage}{8.5cm}
\center
\includegraphics[width=\textwidth]{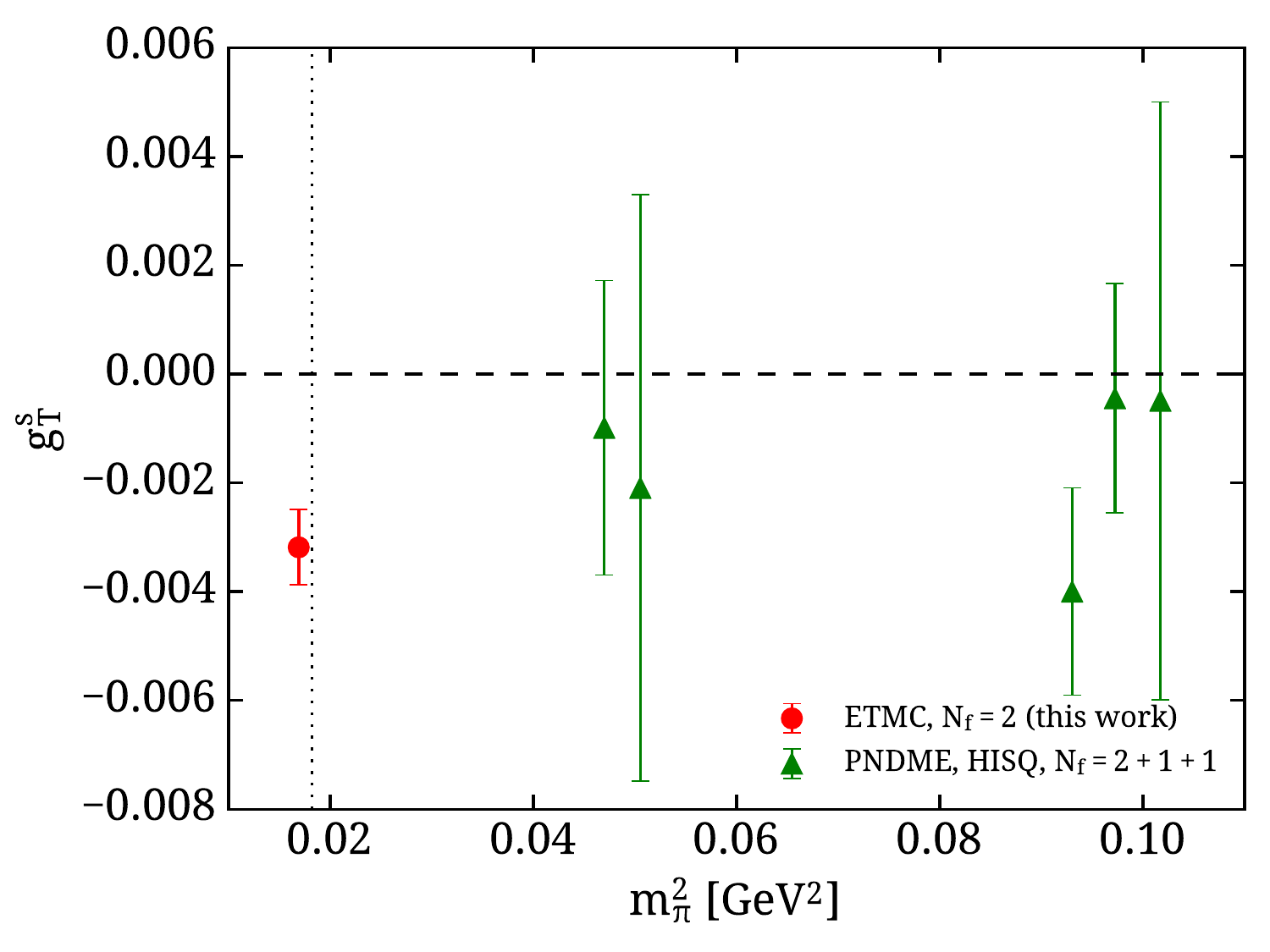}
\end{minipage}
\caption{Comparison of our results using the physical ensemble (red circles) for the strange charges $g_S^s$ (left) and $g_T^s$ (right) with the results from the PNDME collaboration, using $N_f=2+1+1$ staggered fermions (green triangles) from Ref.~\cite{Bhattacharya:2015gma} for $g_S^s$ and Ref.~\cite{Bhattacharya:2015wna} for $g_T^s$.}
\label{fig:gS_gT_strange_world}
\end{figure}

Besides our computation, only the  PNDME  has evaluated the disconnected contributions~\cite{Bhattacharya:2015gma,Bhattacharya:2015wna} at pion masses around $m_\pi=220$~MeV and $m_\pi=310$~MeV.  Disconnected 
contributions, besides the physical ensemble, were also computed for a gauge ensemble of $N_f=2+1+1$ twisted mass fermions  at $m_\pi=373$~MeV~\cite{Abdel-Rehim:2013wlz}. We compare our results for the scalar and tensor charges regarding the disconnected isoscalar as well as the strange contributions in Figs.~\ref{fig:gS_gT_IS_disc_world} and~\ref{fig:gS_gT_strange_world}, respectively.
Concerning the scalar charges, both ETMC and PNDME find that the disconnected contribution to  $g_S^{u+d}$ and $g_S^s$ are non-zero and positive. For $g_S^s$ the PNDME values tend to be larger compared to our values and further study of the systematics is called for. Concerning the tensor charges, our result for the disconnected contribution to $g_T^{u+d}$ is clearly non-zero and negative. While PNDME also find a negative value their error on the value obtained at the smallest  pion mass that they analyzed is much larger making it consistent with zero. We note that the preliminary value of Ref.~\cite{Gambhir:2016jul} agrees with our result. Similarly the value of $g_T^s$ is clearly nonzero and negative.

%As can be seen, both ETMC and PNDME obtain results for the disconnected part of $g_T^{u+d}$ that are consistent with zero. On the other hand, the disconnected contribution to  $g_S^{u+d}$ is found by both to be non-zero and positive. The same is true for $g_S^s$ and $g_T^s$.  PNDME finds  larger values for $g_S^s$ by about two standard deviations at $m_\pi=220$~MeV. It would be interesting for other collaborations to compute $g_S^s$ directly at the physical point in order to have a direct comparison. The ETMC results concerning the disconnected $g_S^{u+d}$ and $g_S^s$ using the $N_f=2+1+1$ ensemble at $m_\pi=373$~MeV~\cite{Abdel-Rehim:2013wlz} are obtained at $t_s=1.65$~fm which is compatible with the separation taken at the physical ensemble from this study, however, the lower values from the former work corroborate the fact that the scalar matrix element is severely contaminated from excited state effects, that tend to decrease its value.
%
% Disconnected Isoscalar charges, world
%
%
% Strange charges, world

%==========================================================================
%==========================================================================

\section{Conclusions}\label{sec:conclusions}

The nucleon scalar and tensor charges are computed within lattice QCD using simulations generated with two dynamical degenerate light quarks with mass fixed to reproduce approximately the physical pion mass. Both isoscalar and isovector combinations are obtained including the disconnected contributions. We also compute the nucleon strange and charm scalar and tensor charges for the first time. After a careful investigation of excited states contributions we obtain in the $\overline{\rm MS}$ at 2~GeV the following values 

\be
g_S^u = 5.200(419)(149)(124)\,,\; g_S^d = 4.270(256)(149)(124)\,,\; g_S^s=0.329(68)(12)(36)\,,\; g_S^c=0.062(13)(3)(5)\;,
\ee
\be
g_T^u = 0.782(16)(2)(13)\,,\; g_T^d = -0.219(10)(2)(13)\,,\; g_T^s=-0.00319(69)(2)(22)\,,\; g_T^c=-0.00263(269)(2)(37)\;,
\ee
where the first error is the statistical error, the second is the systematic error due to the determination of the renormalization functions and the third error is the systematic error due to the excited states, estimated by taking the difference between the mean value obtained from the plateau and two-state fit methods. We stress that both isovector and isoscalar charges are renormalized non-perturbatively with the non-singlet and singlet
renormalization functions, respectively. 
We find that the disconnected contributions to the tensor charge are negligible whereas for the scalar they make about 15\% of the total value in the case of the isoscalar combination. In addition, excited states are found to be more severe in the case of the scalar as compared to the tensor. We note that since these results were produced using one ensemble of twisted mass clover-improved fermions we cannot provide systematics errors due to finite lattice spacing. 
Results from other lattice QCD groups close to the physical point are only reported for the isovector and the connected isoscalar combinations. Overall lattice QCD results are in agreement with a couple of exceptions and produce non-zero values for the scalar strange and charm charges, whereas the strange and charm tensor charge are consistent with zero.

%==========================================================================
%==========================================================================

\section*{Acknowledgments}
We would like to thank the members of the ETMC for a most enjoyable collaboration. We acknowledge funding from the European Union’s Horizon 2020 research and innovation programme under the Marie Sklodowska-Curie grant agreement No 642069.
This work was partly  supported by a grant from the Swiss National Supercomputing Centre (CSCS) under project IDs \texttt{s540} and \texttt{s625} on the Piz Daint system, by a Gauss allocation on SuperMUC with ID \texttt{44060} and in addition used computational resources  from the John von Neumann-Institute for Computing on the Jureca and
the BlueGene/Q Juqueen systems at the research center in J\"ulich. We also acknowledge PRACE for awarding us access to the Tier-0 computing resources Curie, Fermi and
SuperMUC based in CEA, France, Cineca, Italy and LRZ, Germany, respectively. 
%This work was supported, in part, by a grant from the Swiss National Supercomputing Centre (CSCS) under project ID s540. 
We thank the staff members at all sites for their kind and sustained
support.  K.H. and Ch. K. acknowledge support from the Cyprus Research Promotion Foundation under contract T$\Pi$E/$\Pi\Lambda$HPO/0311(BIE)/09.

%=============================================================
%=============================================================
%====== BIBLIOGRAPHY
\clearpage
\bibliographystyle{naturemag}              % Style for bibliogrpahy
\bibliography{references}

%=============================================================
%=============================================================
%====== A P P E N D I C E S

%==============================================================

\end{document}